\documentclass[11pt,a4paper]{article}
\pdfoutput=1
\usepackage{jheppub}                     

\usepackage{amsmath}
\usepackage{amssymb}
\usepackage{graphicx}
 \usepackage{color}
\usepackage[T1]{fontenc} 
\usepackage{slashed}
\usepackage{ulem}
\def\bpm{\begin{pmatrix}} 
\def\epm{\end{pmatrix}} 
\def\bea{\begin{eqnarray}}
\def\eea{\end{eqnarray}}

\definecolor{mkgreen}{rgb}{0.2,.70,.3}

\newcommand{\eq}[1]{eq.~(\ref{#1})}
\newcommand{\fig}[1]{figure~\ref{#1}}
\newcommand{\Fig}[1]{Figure~\ref{#1}}

\newcommand{\REF}[1]{ref.~\cite{#1}}
\newcommand{\SEC}[1]{section~\ref{#1}}
\newcommand{\TAB}[1]{table~\ref{#1}}

\def\SPHENO{{\tt SPheno}}
\def\SARAH{{\tt SARAH}}
\def\MICROMEGAS{{\tt micrOMEGAS 4.1}}

\def\pd[#1]{\frac{\partial}{\partial #1}}
\def\pdd[#1,#2]{\frac{\partial #1}{\partial #2}}

\def\TeV{\mathrm{TeV}}
\def\GeV{\mathrm{GeV}}

\def\hE{\hat{E}}
\def\hEt{\hat{\tilde{E}}}
\def\nn{\nonumber}
\def\ov{\overline}

\def\beq{\begin{equation}}
\def\eeq{\end{equation}}
\def\bal{\begin{align}}
\def\eal{\end{align}}

\def\twovec[#1,#2]{\left( \begin{array}{c} #1 \\ #2\end{array}\right)}
\def\threevec[#1,#2,#3]{\left( \begin{array}{c} #1 \\ #2 \\ #3 \end{array}\right)}
\def\twomatarix[#1,#2][#3,#4]{\left( \begin{array}{cc} #1 & #2 \\ #3 & #4 \end{array}
\right)}
\def\threematrix[#1,#2,#3][#4,#5,#6][#7,#8,#9]{\left( \begin{array}{ccc} #1 & #2 &#3 \\
#4 & #5 & #6\\#7&#8&#9\end{array} \right)}
\def\threediag[#1,#2,#3]{\left( \begin{array}{ccc} #1 & 0 & 0\\
0 & #2 & 0\\0& 0&#3\end{array} \right)}

\def\gsim{\raise0.3ex\hbox{$\;>$\kern-0.75em\raise-1.1ex\hbox{$\sim\;$}}}
\def\lsim{\raise0.3ex\hbox{$\;<$\kern-0.75em\raise-1.1ex\hbox{$\sim\;$}}}

\allowdisplaybreaks

\begin{document}

\title{Dark matter scenarios in a constrained model with Dirac gauginos}

\author[a,b]{M.~D.~Goodsell} 
\author[c]{M.~E.~Krauss}
\author[c]{T.~M\"uller} 
\author[c]{W.~Porod} 
\author[d]{F.~Staub} 

\affiliation[a]{Sorbonne Universit\'es, UPMC Univ Paris 06, UMR 7589, LPTHE, F-75005, Paris, France }
\affiliation[b]{CNRS, UMR 7589, LPTHE, F-75005, Paris, France}
\affiliation[c]{Institut f\"ur Theoretische Physik und Astronomie, 
Universit\"at W\"urzburg\\
97074 W\"urzburg, Germany}
\affiliation[d]{Theory Division, Physics Department, CERN, CH-1211 Geneva 23, Switzerland}

\emailAdd{goodsell@lpthe.jussieu.fr}
\emailAdd{manuel.krauss@physik.uni-wuerzburg.de}
\emailAdd{tobias.mueller@stud-mail.uni-wuerzburg.de}
\emailAdd{porod@physik.uni-wuerzburg.de}
\emailAdd{florian.staub@cern.ch}

\abstract{
We perform the first analysis of Dark Matter scenarios in a constrained model with Dirac Gauginos. The model under investigation is the Constrained Minimal Dirac Gaugino Supersymmetric Standard model (CMDGSSM) where the Majorana mass terms of gauginos vanish. However, $R$-symmetry is broken in the Higgs sector by an explicit and/or effective $B_\mu$-term. This causes a mass splitting between Dirac states in the fermion sector and the neutralinos, which provide the dark matter candidate, become pseudo-Dirac states. We discuss two scenarios: the universal case with all scalar masses unified at the GUT scale, and the case with non-universal Higgs soft-terms. We identify different regions in the parameter space which fulfill all constraints from the dark matter abundance, the limits from SUSY and direct dark matter searches and the Higgs mass. Most of these points can be tested with the next generation of direct dark matter detection experiments. 
}

\preprint{CERN-PH-TH-2015-152}

\maketitle


\section{Introduction}

With the discovery of a particle strongly resembling the Higgs boson we have a wealth of new electroweak precision observables, chief among these being the new scalar's mass. It has therefore become an important endeavour to use this data to constrain models where this quantity can be calculated as much as possible, in particular since there has so far been no clear detection of other new particles at the Large Hadron Collider (LHC). The most important framework where the Higgs boson mass can be calculated from top-down considerations is supersymmetry (SUSY) and, particularly in light of the non-observation of light superpartners, it has become increasingly important to consider more general models beyond the Minimal Supersymmetric Standard Model (MSSM). 
 
One particularly interesting extension of the MSSM is to add Dirac masses for the gauginos\cite{fayet,Polchinski:1982an,Hall:1990hq,fnw,Nelson:2002ca,Antoniadis:2005em,Antoniadis:2006uj,kpw,%
Amigo:2008rc,Plehn:2008ae,Benakli:2008pg,Belanger:2009wf,Benakli:2009mk,Choi:2009ue,Benakli:2010gi,Choi:2010gc,%
Carpenter:2010as,Kribs:2010md,Abel:2011dc,Davies:2011mp,Benakli:2011vb,Benakli:2011kz,Kalinowski:2011zz,Frugiuele:2011mh,%
Itoyama:2011zi,Rehermann:2011ax,Bertuzzo:2012su,Davies:2012vu,Argurio:2012cd,Fok:2012fb,Argurio:2012bi,Frugiuele:2012pe,%
Frugiuele:2012kp,Benakli:2012cy,Itoyama:2013sn,Chakraborty:2013gea,Csaki:2013fla,Itoyama:2013vxa,Beauchesne:2014pra,%
Benakli:2014daa,%
Bertuzzo:2014bwa,Benakli:2014cia,Goodsell:2014dia,Busbridge:2014sha,Chakraborty:2014sda,Ding:2015wma,Alves:2015kia,Alves:2015bba,Martin:2015eca}. This has a number of motivations from both the top down -- such as allowing an (approximate or exact up to gravitational corrections) R-symmetry; simpler models of supersymmetry breaking; permitting $N=2$ supersymmetric subsectors of the theory at high energies -- and bottom up: they allow increased naturalness through supersoftness \cite{fnw,Arvanitaki:2013yja}, contain new couplings that can enhance the Higgs mass, can weaken both LHC search bounds \cite{Heikinheimo:2011fk,Kribs:2012gx,Alves:2013wra} and flavour constaints \cite{kpw, Fok:2012me, Dudas:2013gga}. There is thus a growing literature on their phenomenology (some of which is cited above) to which we refer the reader. 

In previous work \cite{Benakli:2014cia}, (some of) the present authors introduced a constrained unified scenario where, from just a few parameters determined at a Grand Unification (GUT) scale $M_{GUT}$, the entire spectrum of superpartners and the Higgs boson mass could be determined. Preliminary scans determined the general features of the model. Here we shall extend this work by studying the consequences for dark matter under the assumption that the universe has a standard thermal history (i.e. no late-time reheating etc). Since the incontrovertible existence of dark matter is our best direct indication of physics beyond the Standard Model, no new scenario is complete without such an analysis. Interestingly, it can be used as an additional constraint to correlate with other searches (although given our ignorance of the thermal history of the universe it is more difficult to provide definitive exclusions). In addition we will employ a more accurate calculation of the Higgs mass (up to two loop order) by using the latest version of \SARAH\ and the results of \cite{Goodsell:2014bna,Goodsell:2015ira}.

To introduce Dirac gaugino masses we must add adjoint chiral multiplets for each gauge group. In going beyond the MSSM, we are still therefore left with the choice of how much additional matter to include, if any. One motivation is that the gauge couplings should still predict perturbative unification without the need to fine-tune the masses of additional heavy states; in \cite{Benakli:2014cia} it was found that, once two-loop corrections are taken into account, the best choice is perhaps the simplest: we add an extra vector-like doublet of states with the same gauge quantum numbers as the Higgs doublets, and two additional pairs of vectorlike multiplets charged only under hypercharge with charges $\pm 1$. We are then left with the choice of top-down motivation for the supersymmetric and supersymmetry-breaking parameters; broadly we have the choice:
\begin{enumerate}
\item An exact R-symmetry (up to eventual breaking by negligible gravitational effects). The model then resembles the MRSSM with the additional vector-like electrons. 
\item Approximate R-symmetry. 
\end{enumerate} 
We shall take the latter choice as in \cite{Benakli:2014cia}, for several reasons:
\begin{itemize}
\item The R-symmetry must be broken in nature, and, since it is a chiral symmetry, it is reasonable to expect that it is broken in the Higgs sector.
\item We can then write new Higgs couplings which enhance its mass at tree level. In contrast, in the MRSSM the one-loop corrections due to the new higgsino states tend to decrease its mass. 
\item We can make the choice that the new states which assure unification are instead vectorlike leptons. This both simplifies the structure of the model (assuring that the Higgs phenomenology is almost identical to the more minimal models studied in \cite{Belanger:2009wf,Benakli:2011kz,Goodsell:2012fm,Benakli:2012cy}) and gives new predictions for lepton flavour violation experiments. 
\end{itemize}

The question then remains as to what extent the R-symmetry is approximate, i.e. which R-symmetry violating terms we shall allow. We shall, as in previous work and e.g. \cite{Nelson:2002ca}, make the assumption that the superpotential preserves R-symmetry and the main source of supersymmetry-breaking terms is R-symmetric, but that additional interactions that solve the $\mu/B_\mu$ problem (and couple only to the Higgs) do not, and thus we allow a $B_\mu$ term\footnote{We shall later also consider for phenomenological reasons the possibility of a supersymmetric singlet mass term.}. 

With the above assumptions, we can construct a complete phenomenological scenario. In the next section, after a review of our particle content and notation, we will describe the important features of the low energy spectrum based on the high-energy assumptions. In section \ref{SEC:ANALYTIC} we describe the generic predictions for dark matter density and direct detection that we expect. We shall find that, since the model predicts a pseudo-Dirac dark matter candidate, we alleviate the tension between finding the correct dark matter density and the lack of signals for direct detection so far compared to a simple Majorana neutralino. In section \ref{SEC:NUMERICS} we present the results of numerical scans which provide quantitative predictions.


\section{The model}
\label{SEC:MODEL}

Our theory, as defined in \cite{Benakli:2014cia} (to which we refer the reader for a more complete disucssion), consists of the MSSM field content $Q, L, E, U, D, H_u, H_d$ augmented by adjoint superfields for hypercharge, $SU(2)$ and $SU(3)$ denoted $S, T, O$, and new unification fields with representations 
\begin{equation}
\begin{array}{|c|c|} \hline 
\mathrm{Field} & (SU(3), SU(2))_{Y} \\ \hline
R_u & (\mathbf{1}, \mathbf{2})_{-1/2} \\
R_d & (\mathbf{1}, \mathbf{2})_{1/2}\\ \hline
\hE_{1,2} & (\mathbf{1}, \mathbf{1})_1 \\
\hEt_{1,2} & (\mathbf{1}, \mathbf{1})_{-1} \\ \hline
\end{array}
\label{eq:newstates}
\end{equation}
The superpotential of our theory  is\footnote{We use throughout this paper the conventions of Ref.~\cite{Benakli:2014cia}. Note that the definition of the electoweak part of the superpotential differs from that in \cite{Belanger:2009wf} and, furthermore, from the model file implemented in \SARAH.
To use all given numbers as input for  \SARAH/\SPHENO, the following shifts must be applied
\begin{align*}
\mu \rightarrow -\mu, 
 \qquad  \lambda_T \rightarrow \sqrt{2}\,\lambda_T.
\end{align*} 
}
\begin{align}
W =& Y_u^{ij} U_i Q_j H_u - Y_d^{ij} D_i Q_j H_d - Y_e^{ij} E_i L_j H_d \nn\\ 
& +(\mu + \lambda_{S} S ) H_d H_u + 2\lambda_T H_d T H_u \nn\\
&+ (\mu_R + \lambda_{SR} S )R_u R_d + 2\lambda_{TR} R_u T R_d + (\mu_{\hE\, ij} + \lambda_{S\hat{E}\, ij}S)\hE_i \hEt_j\nn\\
&+ \lambda_{SLRi} S L_i R_d + 2\lambda_{TLRi} L_i T R_d + \lambda_{SEij}S E_i \hEt_j\nn\\
& - Y_{\hE i} R_u H_d \hE_i - Y_{\hEt i} R_d H_u \hEt_i  \nn\\
& - Y_{LFV}^{ij} L_i \cdot H_d \hE_j - Y_{EFV}^{j} R_u H_d E_j .
\label{EQ:Superpotential}\end{align}
We retain only R-symmetry preserving soft terms, in particular the Dirac gaugino masses $m_{1D}, m_{2D}, m_{3D}$ and adjoint soft masses $m_S, m_T, m_O$, \emph{except} for a small $B_\mu$ term, assumed to come from a gravitational sector in line with the discussion in the introduction. We expect that in explicit models other small R-symmetry breaking terms may also be generated but that these will not affect the phenomenology; furthermore the inclusion of a $B_\mu$ term is renormalisation running consistent in that it does not lead to the generation of other classes of soft terms\footnote{It is obvious on dimensional grounds that $B$-terms cannot generate $A$-terms or Majorana masses on renormalisation group running, and they do not enter in the RGEs of $m^2$ terms (see e.g. the generic expressions of the RGEs given in Ref.~\cite{Martin:1993zk,Goodsell:2012fm}). The only other RGEs that they could enter in, therefore, are for other $B$-terms  and the scalar tadpole. Now, if we imagine $B_\mu$ to be a background field which transforms under R, if we do not include any explicitly R-symmetry-breaking terms in the superpotential (such as a $S^3$ term) which would also therefore transform under R, then $B_\mu$ cannot appear in any RGE for an R-symmetry-preserving quantity. In our theory this is true for $B_S, B_T, B_O$ and also the scalar tadpole -- so their
 RGEs do not depend on $B_\mu$ to any order. We can also see this clearly at two loops in the explicit expressions given in \cite{Goodsell:2012fm}. On the other hand, other $B$-terms could be generated if they would also violate R; in our theory this is the case for $B_{\hE_i \hEt_j}$ and $B_{E_i \hEt_j}$. However, other than the fact that these are of very little phenomenological importance, their presence in the RGEs must be proportional to the couplings $\lambda_{S\hat{E}\, ij}, \lambda_{SEij} $ which we are neglecting in this analysis.}.

At the GUT scale, we set unified values for the gaugino masses to be $m_{D0}$ and for scalars $m_0$, with the exception of adjoint scalars -- and, in the following, we shall consider scenarios with both universal and non-universal Higgs masses. The adjoint scalars and the Higgs may have direct couplings to the supersymmetry-breaking mediation explaining the non-universality; however the GUT symmetry will enforce the triplet and octet scalars to have a unified mass at the GUT scale $m_\Sigma$ (which may be different from the singlet $m_S$ depending on the completion). Regarding the holomorphic mass terms for the adjoint scalars $B_S, B_T, B_O$ (corresponding to lagrangian terms $\mathcal{L} \supset -\frac{1}{2} B_S S^2 + h.c.$ and similarly for the triplet and octet) we take their values to be parameters that we mostly set to zero except where explicitly stated otherwise. Since we are considering a gravity-mediation-inspired scenario (with potentially mixed F- and D-terms, see e.g. \cite{Kribs:2010md}) the generation of these terms will have a different origin to the Dirac gaugino mass (unlike in typical gauge-mediation scenarios) and so we cannot therefore make any strong claim about what relative size to expect for these terms. However, it is not unreasonable for them to be small in this context as there is no reason to expect a Giudice-Masiero-type term to generate them from F-terms.

In  \cite{Benakli:2014cia} several predictions for the low energy spectrum based on these assumptions were found:
\begin{itemize}
\item Unification takes place at $(1.8 \pm 0.4) \times 10^{17}$ GeV.
\item We have a compressed pattern of soft masses (with deviations of a few percent upon varying the input parameters): 
\begin{align}
&m_{U 33}^2 : m_{Q 33}^2 : m_{Q 11}^2 : m_{D ii }^2 : m_{E ii}^2 : m_{U 11}^2 : m_{L ii}^2  \nn\\
=&0.16 : 0.39 : 0.77 : 0.79 : 0.83 : 0.93 : 1.02\nn
\end{align}
where the ratios are normalised with respect to the common mass at the GUT scale $m_0$. 
\item Sleptons are heavy and quasi-degenerate with the first two generations of squarks. This is because the Dirac gaugino masses do not enter the squark RGEs. 
\item The gaugino masses are in the ratio 
$$m_{1D}/m_{D0} : m_{2D}/m_{D0} : m_{3D}/m_{D0} =  0.22: 0.9 : 3.5.$$
\end{itemize}

In this section we shall outline the anticipated impact of these conclusions for a neutralino dark matter particle in combination with the observed Higgs mass; in the following sections we will perform a detailed numerical study.

\subsection{Electroweak scalar sector}

Introducting the notation 
\begin{eqnarray}
\tilde{m}_{SR}^2&=&\tilde{m}_{S}^2 + B_S + 4m_{1D}^2
\nn\\
\tilde{m}_{TR}^2 &=& \tilde{m}_{T}^2 + B_T + 4m_{2D}^2 
 \end{eqnarray}
the mass matrix for the CP even scalars in the basis $\{h, H,
S_R,T^0_R\}$ is:
\begin{eqnarray}
\label{eq:MassScalar}
\left(\begin{array}{c c c c }
M_Z^2+\Delta_h s_{2\beta}^2 & \Delta_h s_{2\beta}  c_{2\beta}  & \Delta_{hs}    
& 
\Delta_{ht} \\
\Delta_h s_{2\beta}  c_{2\beta} & M_A^2-\Delta_h s_{2\beta}^2   & \Delta_{Hs}    
&\Delta_{Ht}    \\
 \Delta_{hs}     & \Delta_{Hs}      & \tilde{m}_{SR}^2 +\lambda_S^2 \,  \frac{v^2}{2} &  \lambda_S \lambda_T
\frac{v^2}{2} \\
  \Delta_{ht}     &\Delta_{Ht}    &  \lambda_S \lambda_T \frac{v^2}{2} & 
\tilde{m}_{TR}^2 +\lambda_T^2  \frac{v^2}{2} \\
\end{array}\right) 
\end{eqnarray}
where we have defined:
\begin{eqnarray}
\Delta_h&=&\frac{v^2}{2}(\lambda_S^2+\lambda_T^2)-M_Z^2 
\end{eqnarray}
which vanishes when $\lambda_S$ and $\lambda_T$ take their $N=2$ values 
\cite{Antoniadis:2006uj}.
We denote non-diagonal elements describing the mixing of $S_R$ and $T^0_R$ states 
with the light Higgs $h$:
\begin{eqnarray}
 \Delta_{hs} =v[ \sqrt{2} \lambda_S \tilde{\mu} - g_Y m_{1D} c_{2\beta} ] \qquad  \Delta_{ht}=v[ -\sqrt{2} \lambda_T \tilde{\mu} + g_2 m_{2D} c_{2\beta} ] 
\label{EQ:hsht}\end{eqnarray}
where
\begin{align}
\tilde{\mu} \equiv& \mu + \frac{\lambda_S}{\sqrt{2}} v_S - \frac{\lambda_T}{\sqrt{2}} v_T \,,
\label{EQ:mutilde}
\end{align}
while
\begin{eqnarray}
 \Delta_{Hs} = g_Y m_{1D} v  s_{2\beta} ,\qquad
 \Delta_{Ht}  =  - g_2 m_{2D} v  s_{2\beta}  
 \end{eqnarray}
stand for the corresponding mixing with heavier Higgs, $H$. The minimisation conditions for the potential give
\begin{align}
v_S =   - \frac{1}{\tilde{m}_{SR}} \bigg[ t_{S_R}  + \frac{v}{2} \Delta_{hs} \bigg],\qquad
v_T = - \frac{1}{\tilde{m}_{TR}} \bigg[ t_{T^0_R}  + \frac{v}{2} \Delta_{ht} \bigg]
\label{EQ:vsvt}\end{align}
where we have included tadpoles
\begin{equation}
t_{S_R} \equiv - \frac{\partial \Delta V}{\partial S_R}, \qquad  t_{T^0_R} \equiv - \frac{\partial \Delta V}{\partial T_R^0}.
\end{equation}

The significance of the above is that the mixing between the singlet/triplet scalars and the light Higgs $h$ is proportional to the singlet/triplet expectation values. Hence by reducing the mixing term we both allow a larger light Higgs mass and smaller expectation values for the adjoint scalars. While the former is obviously desirable, the latter is also vitally important for electroweak precision tests because a large triplet expectation value would split the masses of the $W$ and $Z$ from their Standard Model values.  
For instance, using 
$\Delta \rho =(4.2 \pm 2.7) \times 10^{-4} $~\cite{Amsler:2008zzb,Aaltonen:2012bp,Abazov:2012bv,Group:2012gb,Barger:2012hr},  
we require
\begin{equation}
\Delta \rho  \simeq 4 \frac{v_T^2}{v^2} <  1 \times 10^{-3} \ (95 \%)
\end{equation}
which is satisfied for $v_T\lesssim 4$ GeV. From equations (\ref{EQ:hsht}) and (\ref{EQ:vsvt}) we see that 
 for large $\tan \beta$, small $\mu$ and/or $\lambda_T$ we have
\begin{align}
0.4 \gtrsim & \left( \frac{m_{2D}}{\tilde{m}_{TR}} \right) \left(\frac{2\ \TeV}{\tilde{m}_{TR}} \right) .
\label{EQ:vtconstraint}
\end{align}
If we were to take the supersoft limit of $\tilde{m}_{TR} = 2 m_{2D} $ this would require $5$ TeV Winos. On the other hand, since $c_{2\beta} < 0$ this implies that we may have a partial cancellation of the mixing terms $\Delta_{hs}$ and $\Delta_{ht}$ if $\mu$ or $\lambda_T/\lambda_S$ are substantial and negative. By making this choice we both enhance the Higgs mass and reduce the shift in the $\rho$ parameter. In the following we shall take $\mu$ negative and all other quantities positive  which allows for a wider parameter space without tuning.

\begin{figure}[htbp]
\centering
\includegraphics[width=.49\linewidth]{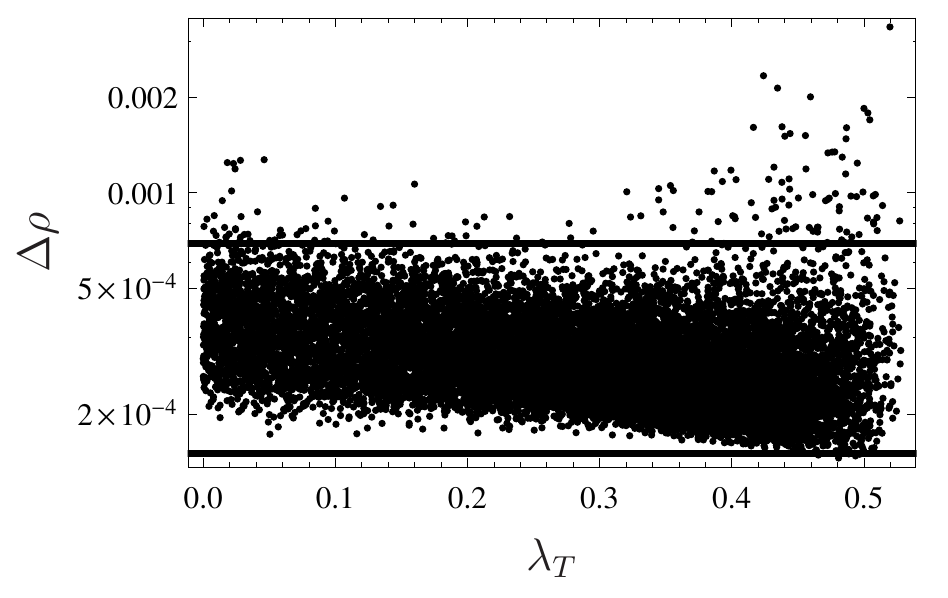}
\includegraphics[width=.49\linewidth]{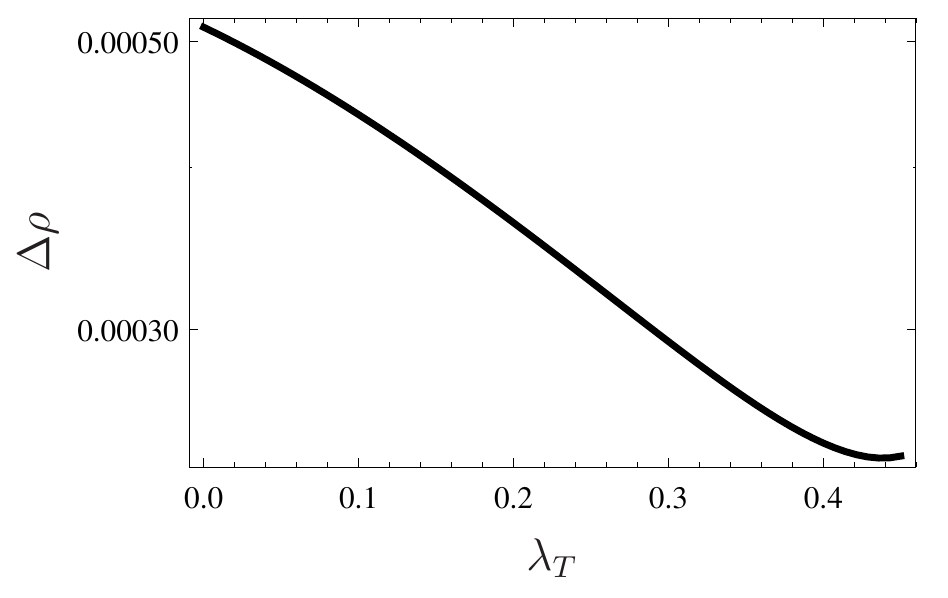}
\caption{
Left: $\Delta \rho$ for data from figure \ref{fig:nonuni:random_relic}. All parameter points with a suitable relic density and right Higgs mass fit the current experimental bounds of $\Delta \rho =(4.2 \pm 2.7) \times 10^{-4} $~\cite{Amsler:2008zzb,Aaltonen:2012bp,Abazov:2012bv,Group:2012gb,Barger:2012hr} (black lines).
Right: Dependence of $\Delta \rho$ on $\lambda_T$. Generically, larger $\lambda_T$ leads to smaller values of $v_T$ and subsequently to smaller $\Delta \rho$. The fixed parameters for this plot are: $m_0$ = 2.7 TeV, $m_{D0}$ = 1.1 TeV, $\tan\beta$ = 2.8, $B_\mu = 4*10^5$ GeV$^2$, $m_S$ = 750 GeV, $m_\Sigma$ = 3 TeV, $\lambda_S$ = 0.45 and $\mu = -625~$GeV.}
\label{fig:nonuni:deltarho}
\end{figure}

In \fig{fig:nonuni:deltarho} the value of $\Delta \rho$ is shown for the points of the random scan, as the experimental value could lead to heavy constraints on our calculations. However we find, that most of the parameter points 
all of the ones with the right relic density fall well within the experimental value of $\Delta \rho =(4.2 \pm 2.7) \times 10^{-4} $~\cite{Amsler:2008zzb,Aaltonen:2012bp,Abazov:2012bv,Group:2012gb,Barger:2012hr}.
As discussed above, this is due to a partial cancellation between the contribution of the Wino mass and $\lambda_T$  to $v_T$ and subsequently to $\Delta \rho$.


\subsection{Sfermions and gauginos}

From \cite{Benakli:2014cia} we recall that, at the SUSY scale:
\begin{align}
m_{1D}/m_{D0} : m_{2D}/m_{D0} : m_{3D}/m_{D0} =&  0.22: 0.9 : 3.5, \\
m_{U 33}^2 : m_{Q 33}^2 : m_{Q 11}^2 : m_{D ii }^2 : m_{E ii}^2 : m_{U 11}^2 : m_{L ii}^2 =&0.16 : 0.39 : 0.77 : 0.79 : 0.83 : 0.93 : 1.02
\end{align}
These ratios will have significant consequences for dark matter: let us suppose firstly that we wish to safely avoid constraints on squarks and the gluino by placing 
\begin{align}
m_{\tilde{q}_{1,2}} >& 1.5\ \TeV, \qquad  m_{\tilde{q}_{3}} >0.7\ \TeV \nn\\
m_{\tilde{g}} >& 1.5\ \TeV \nn\\
m_{\chi^\pm} >& 106\ \GeV.
\end{align}
The loop corrections to the gluino mass are of the order of $\mathcal{O}(100)$ GeV, so neglecting them we find
\begin{align}
m_{D0} \gtrsim& 400~ \GeV.
\end{align}
However, this would yield a very light bino which we should bear in mind. 

For the coloured squarks, they receive supersoft corrections of
\begin{equation}
\delta m_{\tilde{q}}^2 \simeq \frac{4 \alpha_3 m_{3D}^2}{3\pi} \log \frac{m_O^2 + 4 m_{3D}^2}{m_{3D}^2} \simeq 0.6 m_{D0}^2 \rightarrow m_{\tilde{q}} \gtrsim 0.8 m_{D0}.
\label{EQ:SUPERSOFT}
\end{equation}
Thus the coloured squark constraints translate into 
\begin{align}
m_0 > 1.7\ \TeV \qquad \mathrm{OR}\qquad m_{D0} > 1.9\ \TeV.
\end{align}
or more generally $m_0 \sqrt{1+0.8 \frac{m_{D0}^2}{m_0^2}} > 1.7\ \TeV.$

\subsection{Adjoint scalars}

So far we ignored the adjoint scalars. If we take the couplings $\lambda_S, \lambda_T$ to be small, then $m_S, m_T$ barely run at all; but 
\begin{align}
m_O^2 \simeq& 0.81 m_\Sigma^2 -0.36 m_0^2.
\end{align}
Without introducing a $B_O$ term, this implies $m_\Sigma \gtrsim 0.7 m_0$. 

Note that we will require a large mass for the triplet adjoint scalar to avoid a large triplet vev. Indeed, unless we take a large value for $\mu$ and $\lambda_T$ to have a partial cancellation (corresponding to some tuning) we require according to equation (\ref{EQ:vtconstraint}) that $m_{\Sigma} \gtrsim \sqrt{ 11.5\ \TeV \times m_{D0}}$; if $m_{D0}= 400$ GeV then we have $m_\Sigma > 2100\ \GeV. $

\subsection{Neutralinos}
\label{sec:model:neutralinos}

As in the MSSM, the natural dark matter candidate for the CMDGSSM is the neutralino and it shall play a central role in the following. In contrast to the MSSM, however, we have two additional neutralinos and so the mass matrix is larger. In the
$(\tilde{S}, \tilde{B}, \tilde{T}^0, \tilde{W}^0, \tilde{H}^0_d, \tilde{H}^0_u)$ basis it is given by
\begin{align}
\mathcal{M}_{\chi^0} = \left(\begin{array}{c c c c c c}
M_S  & m_{1D} & 0     & 0     &  \frac{ \lambda_S }{\sqrt{2}} v  s_\beta &  \frac{ \lambda_S }{\sqrt{2}} v  c_\beta  \\
m_{1D} & 0   & 0     & 0     & - \frac{v g_Y}{2} c_\beta &  \frac{v g_Y}{2} s_\beta  \\
0     & 0     & M_T  & m_{2D} & - \frac{ \lambda_T }{\sqrt{2}} v s_\beta & - \frac{ \lambda_T }{\sqrt{2}} v  c_\beta  \\
0     & 0     & m_{2D} & 0   &  \frac{v g_2}{2} c_\beta & - \frac{v g_2}{2}  s_\beta  \\
 \frac{ \lambda_S }{\sqrt{2}} v  s_\beta &  - \frac{v g_Y}{2} c_\beta &- \frac{ \lambda_T }{\sqrt{2}} v s_\beta  &  \frac{v g_2}{2} c_\beta  & 0    & \tilde{\mu} \\
\frac{ \lambda_S }{\sqrt{2}} v  c_\beta  & \frac{v g_Y}{2} s_\beta & - \frac{ \lambda_T }{\sqrt{2}} v  c_\beta  &  - \frac{v g_2}{2}  s_\beta & \tilde{\mu} & 0    \\
\end{array}\right) \,.
\label{EQ:NeutralinoMass}\end{align}
We have included supersymmetric masses $M_S, M_T$ in the above for later convenience, but in the pure CMDGSSM these are both zero. 

Clearly, since the ratio of the bino to wino masses is so large at low energies, in the CMDGSSM the Wino will be almost decoupled from a dark matter perspective. We can therefore consider the LSP to be either a bino, higgsino or a mixture -- and indeed the mixing will be very important. In the limit that $|\mu| \gg m_{1D}$ or vice versa the LSP will be a \emph{pseudo-Dirac} particle and this has a significant effect on the dark matter density, as we shall see in the following.


\section{Consequences for dark matter}
\label{SEC:ANALYTIC}

In this section we shall examine analytically the expectations for the dark matter properties of a neutralino in the CMDGSSM, analysing in turn the pseudo-Dirac bino, higgsino and then a mixed state.

\subsection{Pseudo-Dirac bino}

If $\mu$ is sufficiently large, then the lightest neutralino will be a pseudo-Dirac bino. We can then 
estimate the relic density to be \cite{Hsieh:2007wq}:
\begin{align}
\Omega h^2 =& \frac{16 \pi x_f}{g_Y^4 \sqrt{g_*}}  \frac{8.7 \times 10^{-11} \GeV^{-2}}{m_{1D}^2 \sum \frac{N_f Y_f^4}{M_{\tilde{f}}^4}} \,.
\end{align}
We shall assume that the new vector-like fermions are too heavy for the bino to annihilate to. Then we find, using $x_f = 20$ and $g_* = 96$ that 
\begin{align}
\Omega h^2 \simeq& 0.2 \times \left(\frac{m_0}{2\ \TeV}\right)^2 \frac{m_0^2}{m_{D0}^2}\,.
\end{align}
From the bounds above, we require a large $m_0$ (the minimum values compatible with the above are $m_0 \sim 1.4~\TeV, m_{D0} \sim 1300\ \GeV$ giving a bino of $\sim 290$ GeV) typically implying a rather heavy neutralino. However, this then presents three challenges for the Higgs mass: first, we require large $m_0$ to give a significant stop contribution, while we then require $m_{D0} > m_0$ for the dark matter density -- which poses difficulties for the singlet and triplet neutral scalars mixing; in particular, this will lead to $m_{2D} > m_0$ and we would thus need a very large adjoint scalar mass to prevent erasing the Higgs quartic coupling. In addition, such large values of $m_{D0}$, while still remaining natural, may place into doubt the reliability of the two-loop Higgs mass calculation on a technical level.

If we were to consider $m_{D0} \sim 1300\ \GeV$ then to ensure that $v_T$ is not too large we require $m_\Sigma \gtrsim 3900$ GeV -- although this assumes that $\lambda_T$ is small.

Note that in \cite{Belanger:2009wf} it was found that slepton coannihilation could provide an acceptable relic density -- even with rather different masses for the Dirac bino and the sleptons (unlike the Majorana case). However, this is not reasonable to expect here given the boundary conditions in the UV.

\subsection{Higgsino}

For a sufficiently large value of $m_{D0}$ a higgsino could be the dark matter candidate. For $m_{1D} \gg \mu$ the eigenstates would be split via mixing with the bino by an amount \cite{Belanger:2009wf} $\Delta \chi \sim \mu \frac{M_Z^2 s_W^2}{m_{1D}^2}$; for the values of interest this is \emph{always} small enough to allow co-annihilation during freezeout, but much too large to allow for inelastic dark matter. Hence we can estimate the relic density using the conventional approximation for higgsino dark matter of \cite{ArkaniHamed:2006mb}:
\begin{align}
\Omega h^2 \simeq 0.1 \left( \frac{\mu}{\TeV} \right)^2 \,.
\label{eq:analytics:higgsino_dm}
\end{align}  
However, such a large value of $\mu$ requiring an even larger value of $m_{1D}$ implies that the gluino mass would be greater than $16$ TeV; while this could be considered it is beyond the realm of validity of our codes.

\subsection{Mixed bino-higgsino}

This is the most natural option: that $\mu$ is small and therefore the higgsino mixes with the bino. Here we can at first appeal to the results of \cite{Belanger:2009wf}: if the LSP is much lighter than the sfermions then it was concluded that the correct relic density is obtained for a higgsino fraction $f_{h,i} \equiv |N_{i5}|^2 + |N_{i6}|^2$ of about $0.2$. In that work expressions for the mixing were given in the limits $|\mu| \gg m_{1D} $ and $|\mu| \ll m_{1D} $, but these limits are not appropriate for the required higgsino fraction and we therefore give here more accurate expressions. Taking all of the paramters in the neutralino mass matrix (\ref{EQ:NeutralinoMass}) to be real and positive, except for $\mu$ which (as discussed above) we take negative, we define the mass eigenstates in terms of the original eigenstates via $\chi^\prime_i =  N_{ij} \chi_j$ where the usual rotation matrix includes phases to yield the masses to be real and positive, which to leading order has relevant entries given by\footnote{We shall write $|\mu|$ for clarity since we are taking $\mu$ to be negative; strictly speaking we should write $-\mu$.}
\begin{align}
N_{11} \simeq& N_{12} \simeq 1/\sqrt{2}, \quad N_{21} \simeq - i/\sqrt{2}, \quad N_{22} \simeq i/\sqrt{2} \nn\\
N_{15} \simeq& \frac{v c_\beta}{2\sqrt{2}(\mu^2 - m_{1D}^2)} \bigg[m_{1D} ( g_Y -  t_\beta \sqrt{2}\lambda_S) +  |\mu| ( t_\beta g_Y + \sqrt{2} \lambda_S) \bigg] \nn\\
N_{16} \simeq& - \frac{v c_\beta}{2\sqrt{2}(\mu^2 - m_{1D}^2)} \bigg[m_{1D} ( t_\beta g_Y +  \sqrt{2}\lambda_S) +  |\mu| ( g_Y - t_\beta \sqrt{2} \lambda_S) \bigg] \nn\\
N_{25} \simeq& \frac{ i v c_\beta}{2\sqrt{2}(\mu^2 - m_{1D}^2)} \bigg[m_{1D} ( g_Y +  t_\beta \sqrt{2}\lambda_S) -  |\mu| ( t_\beta g_Y - \sqrt{2} \lambda_S) \bigg] \nn\\
N_{26} \simeq&  \frac{i v c_\beta}{2\sqrt{2}(\mu^2 - m_{1D}^2)} \bigg[-m_{1D} ( t_\beta g_Y -  \sqrt{2}\lambda_S) +  |\mu| ( g_Y + t_\beta \sqrt{2} \lambda_S) \bigg]\,.
\end{align}
The two lightest mass eigenstate values are 
\begin{align}
m_{1,2} \simeq& m_{1D} - \frac{v^2}{8(\mu^2 - m_{1D}^2)} \bigg[ 2\sqrt{2} g_Y \lambda_S |\mu| c_{2\beta} + m_{1D} ( 2 \lambda_S^2 + g_Y^2)\bigg] \nn\\
&\pm \frac{v^2 |\mu| s_{2\beta}}{8(\mu^2 - m_{1D}^2)} (2 \lambda_S^2 -g_Y^2  ) \nn\\
\end{align}
where we have defined the masses to be positive, and the upper sign on the second line corresponds to the first eigenvalue. We therefore see that the typical mass splitting for the two lightest eigenstates is $ \mathcal{O} (10)$ GeV and therefore coannihilation is always important in this model. This has significant consequences for the dark matter phenomenology, in that the relic abundance can be reduced but at the same time (as we shall see) the direct detection cross-sections will be suppressed.

From the above we see that the higgsino fractions of the two lightest eigenstates are given by
\begin{align}
f_{h,1} \simeq& \frac{v^2}{16( \mu^2 - m_{1D}^2)^2} \bigg[ (\mu^2 + m_{1D}^2) (g_Y^2 + 2\lambda_S^2) + 2 m_{1D} |\mu| \bigg( 2 \sqrt{2} c_{2\beta} g_Y \lambda_S  + s_{2\beta} (g_Y^2 - 2\lambda_S^2) \bigg)\bigg] \nn\\
f_{h,2} \simeq& \frac{v^2}{16( \mu^2 - m_{1D}^2)^2} \bigg[ (\mu^2 + m_{1D}^2) (g_Y^2 + 2\lambda_S^2) + 2 m_{1D} |\mu| \bigg(2 \sqrt{2} c_{2\beta} g_Y \lambda_S  - s_{2\beta} (g_Y^2 - 2\lambda_S^2) \bigg)\bigg] .
\end{align}
If we take $t_\beta = 1$ then we can write
\begin{align}
f_{h_1} \rightarrow&  \frac{M_Z^2 s_W^2}{4  (|\mu| - m_{1D})^2} + \frac{\lambda_S^2 v^2}{8(|\mu| + m_{1D})^2}, \qquad f_{h_2} \rightarrow  \frac{M_Z^2 s_W^2}{4  (|\mu| + m_{1D})^2} + \frac{\lambda_S^2 v^2}{8(|\mu| - m_{1D})^2}.
\end{align}
Hence we see that as $\lambda_S$ increases the second eigenvalue becomes the lighter and also has a larger higgsino component, while for small $\lambda_S$ it is the other way round. If, for example, we take $\lambda_S = 0.5$ then a higgsino fraction of $0.2$ is possible even for large $|\mu|, m_{1D}$ if $||\mu| - m_{1D}| \simeq 100$ GeV. Thus we see that in the CMDGSSM the higgsino fraction is naturally larger than in the MSSM for a given splitting between bino and higgsino masses, and this will also help to naturally obtain the correct relic abundance. 

Since the sfermion spectrum is heavy, the relevant interactions for the dark matter density and detection are with the higgs and Z, and to a lesser extent with the W and charginos (these latter are relevant for t-channel annihilation processes). Concentrating on the Higgs and Z portals\footnote{By `portals' refer to (as usual) the mediators exchanged in the interactions.}, the interactions can be written as  
\begin{align}
\mathcal{L} \supset \frac{c_{h\chi^0_i \chi^0_j}}{2} h (\chi^0_i \chi^0_j + \ov{\chi}^0_i\ov{\chi}^0_j) + c_{Z\chi^0_i \chi^0_j} \ov{\chi}^0_i\ov{\sigma}^\mu \chi^0_j Z_\mu 
\end{align}
where we can write 
\begin{align}
c_{h\chi^0_i \chi^0_j} = \frac{1}{2} \bigg\{& g_2 ( t_W \ov{N}_{i2} - \ov{N}_{i4})(s_\beta \ov{N}_{j6} - c_\beta \ov{N}_{j5}) \nn\\
&+ \sqrt{2} ( \lambda_S \ov{N}_{i1} - \lambda_T \ov{N}_{i3}) (s_\beta \ov{N}_{j5} + c_\beta \ov{N}_{j6}) + (i \leftrightarrow j) \bigg\} \\
c_{Z\chi^0_i \chi^0_j} =& \frac{e}{2c_W s_W} ( N_{i5} \ov{N}_{j5}-  N_{i6}\ov{N}_{j6}).
\end{align}
In the CMDGSSM the Wino is heavy and so we can neglect $N_{13}, N_{14}$.

\subsubsection{Relic density}

The interactions via the Z portal are spin dependent and complicated. However, from \cite{Arcadi:2014lta} we see that for the mass range of interest to us ($> 100$ GeV) to match the correct relic density via the Z portal alone would require
\begin{align}
| \ov{N}_{i5} N_{j5} - \ov{N}_{i6} N_{j6}| \sim 0.01 \div 0.1
\end{align}
although the model in that reference contained no charginos and so it is not obvious if we can use that result here. 

On the other hand, for a purely Higgs-mediated interaction, we can approximate the relic density as coming from degenerate Majorana fermions while their mass differences are comparable or less than the temperature at freezeout $T_f \sim m_{\chi}/20$. Using Micromegas for a single Majorana fermion and then extrapolating we find for masses above the top production  threshold
\begin{align}
\Omega h^2 \simeq 0.112 \times  \frac{0.07}{\sum_{i,j=1}^2|c_{h\chi^0_i \chi^0_j}|^2 } \times \left( \frac{m_\chi}{200\ \mathrm{GeV}}\right)^2\,.
\label{EQ:higgsportalomega}\end{align}
For larger masses and smaller higgsino fractions we expect this to be the most important process: the annihilation cross-section is roughly proportional to the higgsino fraction, while for the Z portal it is roughly proportional to the higgsino fraction squared. Hence we give 
\begin{align}
\sum_{i,j=1}^2|c_{H\chi^0_i \chi^0_j}|^2 \simeq \frac{v^2}{8 (\mu^2 - m_{1D}^2)^2} \bigg[& m_{1D}^2 ( g_Y^2 + 2 \lambda_S^2)^2  \\
&+4 \sqrt{2}(g_Y^2+ 2 \lambda_S^2) c_{2\beta} g_Y \lambda_S m_{1D} |\mu|   \nn\\
&+ \mu^2 \bigg(  2s_{2\beta}^2 (g_Y^4 + 4 \lambda_S^4) + 8 c_{2\beta}^2 g_Y^2 \lambda_S^2  \bigg)\bigg]. \nn
\label{EQ:chij}\end{align} 
If we take $t_\beta= 1$ and neglect subleading corrections in  $(|\mu| - m_{1D})/v$, we can write
\begin{align}
\Omega h^2 \sim 0.25 \times \frac{ (|\mu| - m_{1D})^2 }{v^2 ( 3g_Y^4 + 4 g_Y^2 \lambda_S^2 + 12 \lambda_S^4)} \times \left( \frac{m_{1D}}{200\ \mathrm{GeV}}\right)^2
\end{align}
so for example for $\lambda_S = 0.5, t_\beta=1$ we find
\begin{align}
\Omega h^2 \sim  0.112 \times \left(\frac{ |\mu| - m_{1D} }{120\ \mathrm{GeV}}\right)^2 \times \left( \frac{m_{1D}}{200\ \mathrm{GeV}}\right)^2 .
\end{align}
However, as we increase $m_{1D}$, we note that the approximation is not good for $|\mu| - m_{1D} \lesssim \lambda_S v s_\beta/\sqrt{2} $. Later on we shall consider $\lambda_S = 0.5$ for which we require $|\mu| - m_{1D} > 80$ GeV, in which case we should evaluate the couplings numerically. Taking $\lambda_S = 0.5, t_\beta = 2.7$ we find for $|\mu| - m_{1D} = 50 $ GeV that $\sum_{i,j=1}^2|c_{H\chi^0_i \chi^0_j}|^2 \simeq 0.07$. For masses somewhat greater than $200$ GeV, however, we find that the mass difference required to enhance the higgsino mixing becomes small enough to potentially bring chargino coannihilation into consideration, and then the above approximation for the relic density will no longer be valid. 

\subsubsection{Direct detection}

Since our dark matter particle is always pseudo-Dirac in nature, only the lightest mass eigenstate survives and may have a rather small direct detection cross-section. Indeed we have spin-dependent and  spin-independent cross-sections \cite{Cheung:2012qy}:
\begin{align}
\sigma_{SI} = 8 \times 10^{-43} \mathrm{cm}^2 \left(c_{h\chi_1^0 \chi_1^0}\right)^2, \qquad \sigma_{SD} = 3 \times 10^{-37} \mathrm{cm}^2 \left( c_{Z\chi_1^0 \chi_1^0} \right)^2 .
\end{align}
In this case to satisfy the  bounds we should have $|c_{h\chi_i^0 \chi_i^0}| < 0.05$ and $|c_{Z\chi_i^0 \chi_i^0}| \lesssim 0.05  $.
In the case of a single neutralino with no coannihilation this would pose a problem to obtain the correct relic density, however in our case these bounds are generically satisfied either because the pseudo-Dirac bino annihilates efficiently through the higgs or Z portals, or because there is additional coannihilation via the Z. We can give approximate formulae for these couplings (for both light eigenvalues, since the lightest state depends on the choice of parameters):
\begin{align}
c_{h\chi_1^0 \chi_1^0} \simeq& -\frac{v}{4(\mu^2 - m_{1D}^2)} \bigg[ m_{1D} (g_Y^2 + 2 \lambda_S^2) + |\mu|\big(  2 \sqrt{2} c_{2\beta} g_Y \lambda_S + s_{2\beta} (g_Y^2 -2 \lambda_S^2)\big)  \bigg]\nn\\
c_{h\chi_2^0 \chi_2^0} \simeq& -\frac{v}{4(\mu^2 - m_{1D}^2)} \bigg[ m_{1D} (g_Y^2 + 2 \lambda_S^2) + |\mu|\big(  2 \sqrt{2} c_{2\beta} g_Y \lambda_S - s_{2\beta} (g_Y^2 -2 \lambda_S^2)\big)  \bigg] \nn\\
c_{Z\chi_1^0 \chi_1^0} \simeq& -\frac{v^2}{16c_W s_W (\mu^2 - m_{1D}^2)} \bigg[ c_{2\beta}   (g_Y^2 -2 \lambda_S^2) - 2 \sqrt{2} s_{2\beta} g_Y \lambda_S \bigg]\nn\\
c_{Z\chi_2^0 \chi_2^0} \simeq& -\frac{v^2}{16c_W s_W (\mu^2 - m_{1D}^2)} \bigg[ c_{2\beta}   (g_Y^2 -2 \lambda_S^2) + 2 \sqrt{2} s_{2\beta} g_Y \lambda_S \bigg].
\end{align}
We see that for larger values of $m_{1D}, \mu$ the spin-dependent scattering will become very small. Taking $t_\beta = 2.71, \lambda_S = 0.5,m_{1D} = 200$ GeV again we find
\begin{align}
c_{h\chi_2^0 \chi_2^0} \simeq& -0.05 \times \left( \frac{100\ \mathrm{GeV}}{|\mu| - m_{1D}}\right), \qquad c_{Z\chi_2^0 \chi_2^0} \simeq -0.01 \times \frac{300\ \mathrm{GeV}}{|\mu|} \times \left( \frac{100\ \mathrm{GeV}}{|\mu| - m_{1D}}\right).
\end{align}
This implies that it is very difficult to satisfy both direct detection constraints and the correct relic density through a pure Higgs mediated interaction involving just a pseudo-Dirac bino; we will have to appeal to the Z portal or resonances.

\subsection{Resonances}

We have seen that imposing that our gluino should be heavier than current simplified model bounds leads to a bino heavier than $90$ GeV and therefore a Standard-Model-Higgs resonance is not possible. Therefore the available resonances are the MSSM-like Heavy higgs and pseudoscalar. At tree level and with small mixing with the singlet these all have masses near
\begin{align}
m_A^2 \simeq \frac{2B_\mu}{s_{2\beta}}
\end{align}
as in the MSSM. This is preserved at loop level even with large $\lambda_S$ (up to $0.7$) as can be seen from the example spectra in \cite{Benakli:2014cia}. The widths of the heavy Higgs and pseudoscalars are rather large -- neglecting QCD corrections the partial width to tops is \cite{Djouadi:2005gi}
\begin{align}
\Gamma (H \rightarrow \ov{t} t) \simeq \frac{3 G_F}{4 \sqrt{2} \pi t_\beta^2 } m_H m_t^2 \left(1-\frac{4m_t^2}{m_H^2}\right)^{3/2} \nn\\
\Gamma (A \rightarrow \ov{t} t) \simeq \frac{3 G_F}{4 \sqrt{2} \pi t_\beta^2 } m_H m_t^2\left(1-\frac{4m_t^2}{m_H^2}\right)^{1/2} 
\end{align}
which are $\mathcal{O}(30)$ GeV for $500$ GeV Heavy Higgses; when we consider that this is comparable to the kinetic energy of the neutralino at freezeout we see that a large tuning is not necessary to sit near the resonance. On the other hand, if we want to consider the decoupling regime where the $m_A \gg m_h$ then we would require gluinos of mass $m_{\tilde{g}} \gtrsim 16 m_A$ which could potentially take us beyond the regime of validity of the codes.


\section{Numerical results}
\label{SEC:NUMERICS}

We have implemented the model in the spectrum generator generator \SARAH\ 
\cite{Staub:2008uz,Staub:2009bi,Staub:2010jh,Staub:2012pb,Staub:2013tta,Staub:2015kfa} as described in detail 
in \REF{Benakli:2014cia}. \SARAH\ generates the routines for a precise spectrum calculation in \SPHENO\ 
\cite{Porod:2003um,Porod:2011nf}. The latter is used to evolve the full RGEs at the two-loop level and to
calculate all pole-masses for supersymmetric 
particles and the corresponding mixing matrices at the one-loop level in the $\overline{\rm DR}$ scheme.
The neutral scalar Higgs masses are calculated at 
two loops using an effective potential approach \cite{Goodsell:2014bna,Goodsell:2015ira} 
which we cross-checked with the diagrammatic two-loop results including the $\alpha_s \alpha_t$ corrections for Dirac gauginos plus the known $\alpha_t^2 +\alpha_t \alpha_b+ (\alpha_b+\alpha_\tau)^2$ contributions from the MSSM.
The spectrum is then passed through the {\tt SLHA+} interface \cite{Belanger:2010st} to \MICROMEGAS\ \cite{Belanger:2001fz,Belanger:2004yn,Belanger:2006is,Belanger:2013oya,Belanger:2014vza} using {\tt CalcHep} \cite{Pukhov:2004ca,Boos:1994xb} model files generated by \SARAH.
\MICROMEGAS\ allows the computation
of the relic dark matter density as well as dark matter-nucleon cross sections  relevant for direct detection experiments.
In \TAB{tab:SMinput} we list the Standard Model input parameters.

In particular we consider two different realizations: 
\begin{enumerate}
\item  {\bf universal scalar masses }: all scalar masses, the sfermions as well as the Higgs bosons,
 are fixed to a common value $m_0$ at the GUT scale, which is determined by the requirement $g_1=g_2$.
 As in the MSSM, the values for $|\mu|$ and $B_{\mu}$ are calculated from the tadpole-equations in the Higgs
 sector. Moreover, also $v_T$ and $v_S$ are calculated from the tadpole-equations using numerical methods.
 For low values of $m^2_0$ the pseudoscalar octet can become tachyonic due to large negative loop corrections.
 Therefore we also allow in general for a non-zero  $B_{O}$ term, which also shifts the mass of the scalar
 octet but  has a negligible effect
 on the rest of the spectrum. Note, that although the superpotential parameter $M_O$ violates 
 $R$-symmetry, the corresponding soft SUSY-breaking bilinear parameter $B_O$ does not.

\item {\bf non-universal Higgs masses (NUHM)}: 
 the sfermions and the scalar fields of $R_{u,d},\hat E_i,\hat{\tilde E}_i$ have a common mass $m_0$
 at the GUT-scale. The parameters $\mu$, $B_\mu$  and $m_{O}^2 = m_T^2 = m^2_\Sigma$ are defined at $M_{GUT}$,
 whereas $m^2_{S}$ is given at the SUSY scale. We choose the value for $m^2_{S}$ at the SUSY scale because
 when defining $m^2_{S}$ at $M_{GUT}$ we often encountered tachyonic states, in particular in scenarios with
  large $\lambda_S$ and $\lambda_T$.  The tadpole equations are solved for  $m_{H_u}^2$ and
  $m_{H_d}^2$,  $v_S$ and $v_T$. This leads in general to soft masses for the Higgs doublets at the GUT scale which are different to those 
  of the other scalars. Moreover, in contrast to the universal case we also define $\lambda_S$ and $\lambda_T$ at 
  the SUSY scale rather than at the GUT scale for a better control over the Higgs mass. 
\end{enumerate}

\begin{table}[t]
\centering
\begin{tabular}{c|c||c|c} \hline \hline
 ~~$\alpha_{em}^{-1}$~~ & 127.929338 & ~~$G_\mu$~~ & $1.16639 \cdot 10^{-5} \text{GeV}^{-2}$ \\
 ~~$\alpha_S$~~ & 0.11720 & ~~$M_Z$~~ & 91.18760 GeV \\
 ~~$m_b(m_b)$~~ & 4.2 GeV & ~~$m_t$~~ & 172.9 GeV \\
 ~~$m_\tau $~~ & 1.777 GeV &  &\\
 \hline 
 \hline
\end{tabular}
\caption{Input values for the Standard Model parameters taken at $M_Z$ unless
  otherwise specified.}
\label{tab:SMinput}
\end{table}

Before we present our results, a few comments and conventions are in order: First, we want to stress that all calculations
assume the standard thermal history of the universe. However, it might well be
that this is changed due to  additional states appearing at the various breaking stages which could lead
for example to additional entropy production which depend on the details of the UV-completion of the model,
see e.g.\
\cite{Kane:2015qea,Allahverdi:2013noa,Staub:2009ww,Arbey:2009gt,Acharya:2008bk,Baltz:2001rq} for related considerations. 
Secondly, these
calculations do not include higher corrections for the annihilation cross sections
which can have quite some impact on the resulting relic density
 \cite{Harz:2014gaa,Boudjema:2011ig,Herrmann:2009wk,Baro:2009na,Freitas:2007sa,Herrmann:2007ku,Baro:2007em}.
Clearly these uncertainties are much larger than the ones from observation which are already at the percent
level. Therefore we allow below for a somewhat larger range for an acceptable dark matter relic density compared 
to the current 2$\sigma$ of $\Omega h^2 = [0.1153,0.1221]$ preferred by experiment \cite{Ade:2013zuv}.

The colour coding used in the scatter plots is always as follows: 
\begin{itemize}
 \item {\bf Red points} have Higgs mass in the range
 \begin{equation}
 \label{eq:ranges}
  121~{\rm{GeV}} < m_h < 129~{\rm{GeV}} \hspace{1cm} \text{and} \hspace{1cm} 0.05 < \Omega h^2 < 0.20
 \end{equation}
\item {\bf Black points} have a Higgs mass outside of this range. 
\item {\bf  Green points}  have a relic density outside of the range but Higgs mass within the correct range. 
\item {\bf Grey points} have a neutralino-nucleon cross-section that has already 
been excluded by spin-dependent and/or spin-independent direct detection measurements by 
LUX \cite{Akerib:2013tjd} and/or XENON100 \cite{Aprile:2013doa,Aprile:2012nq}.
\end{itemize}
The points with {\bf darker shades} of red, green and grey have the same properties
as the brighter red and green as well as black points but with the difference that the associated
spin-independent direct detection cross section is accessible for the next generation of the XENON experiment, XENON1T \cite{XENON:2013}. Hence, dark red points are the ones which correspond to the most promising scenarios because they can be tested in the near future.


\subsection{Universal boundary conditions}
We start with  the more constrained setup with universal Higgs masses. First, as in previous work and discussed in the introduction we assume that the only 
$R$-symmetry breaking parameter in the superpotential is an explicit $B_\mu$-term. In a second
analysis we also allow a non-zero  $M_S$. This purely phenomenological choice allows us to explore different possibilities to
obtain the correct relic density by changing some neutralino masses and mixing entries without 
 significantly changing the masses and properties of the other supersymmetric
particles or the Higgs bosons.

\subsubsection{Case 1: $M_S = 0$}
\label{sec:uni:msneq0}
\begin{table}[hbt]
\centering
\begin{tabular}{l|r c l||l|r c l} \hline \hline
 ~~$m_0$~[GeV] ~~& 400 &...& 3000 & ~~$\lambda_S$~~ & $-1.5$ &...& 1.5  \\
 ~~$m_{D0}$~[GeV]~~ & 400 &...& 3000 & ~~$\lambda_T$~~ & $-1/\sqrt{2}$ &...& $1/\sqrt{2}$ \\
 ~~$\tan \beta$~~ & 1.1 &...& 40 & ~~$B_{O}$~[GeV$^2$]~~ & $-2 \cdot 10^6$ &...& $5 \cdot 10^5$ \\
 \hline 
 \multicolumn{8}{c}{Case 1: scans with $M_S=0$:}\\
 \hline
 ~~$M_S$~~ & &0& & ~~$B_S$~[GeV$^2$]~~ & $-5 \cdot 10^6$ &...& $5 \cdot 10^6$ \\
 \hline 
 \multicolumn{8}{c}{Case 2: scans with $M_S \neq 0$:}\\
 \hline
 ~~$M_S$~[GeV]~~ & $0$ &...& $10^4$ & ~~$B_S$~[GeV$^2$]~~ & $-5 \cdot 10^7$ &...& $5 \cdot 10^7$ \\
 \hline
 \hline
\end{tabular}
\caption{Ranges of the varied parameters using universal GUT-scale boundary conditions. All input parameters
are taken at the GUT scale.}
\label{tab:uni:parameter_ranges}
\end{table}
%
\begin{figure}[t]
\centering
\includegraphics[width=.49\linewidth]{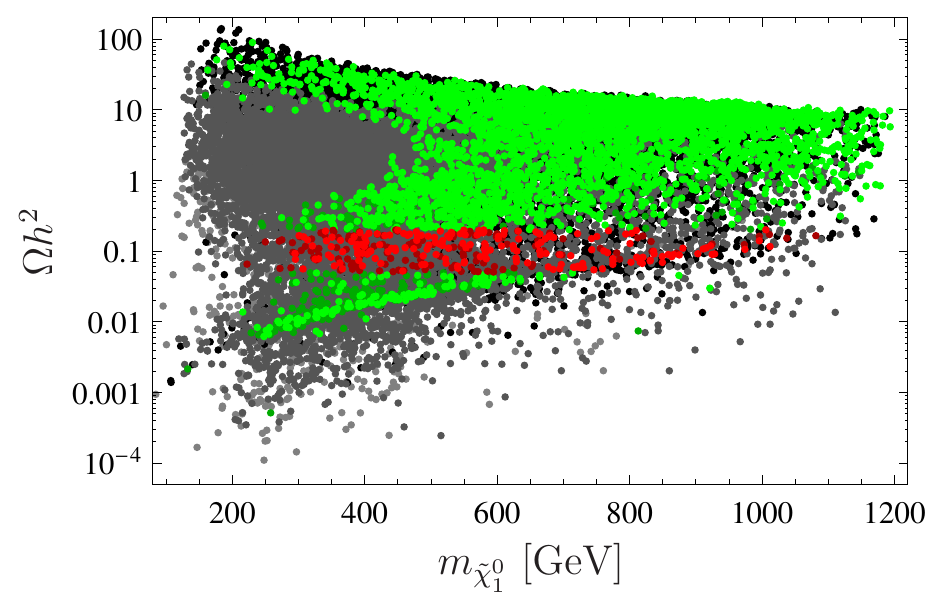}
\includegraphics[width=.49\linewidth]{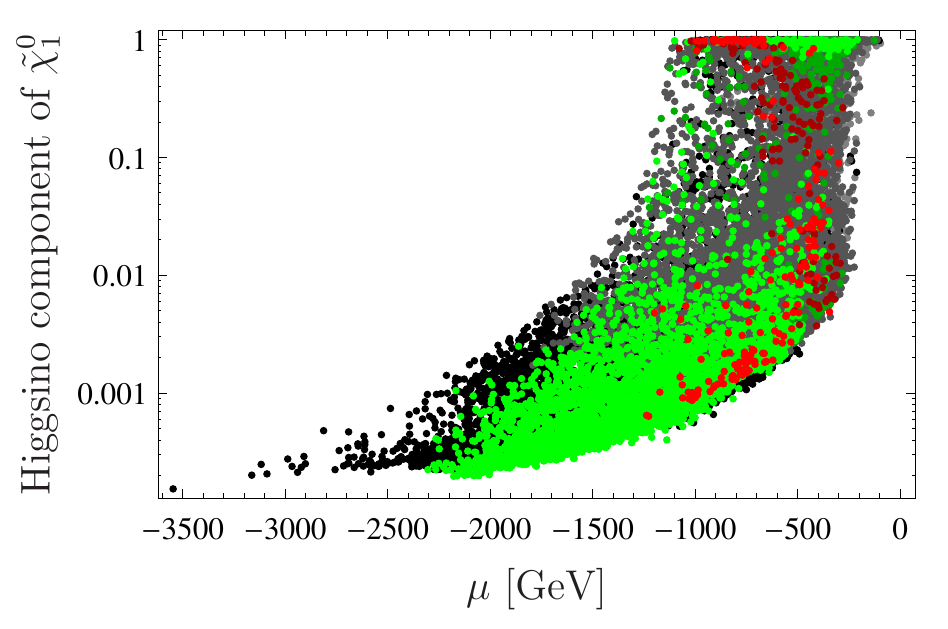}
\caption{Results of a random scan using the 
parameter ranges of \TAB{tab:uni:parameter_ranges}.
The left plot shows the relic density $\Omega h^2$ versus $m_{\tilde \chi^0_1}$ and the right one
gives the  higgsino component of the LSP
$|N_{15}|^2+|N_{16}|^2$
of $\tilde \chi^0_1$
 versus $\mu$. 
The colour code is as described in the text in the context of eq.~(\ref{eq:ranges})}
\label{fig:uni:random_relic}
\end{figure}

For the chosen parameter ranges as given  in \TAB{tab:uni:parameter_ranges}
we find that the tadpole conditions imply $|\mu|	 \gsim 200$~GeV.
Thus, this is approximately the minimal expected mass of a 
higgsino-like LSP. On the other side, we find higgsino-like DM candidates with masses 
up to the TeV range with the correct abundance as can be seen in
the right plot of \fig{fig:uni:random_relic}. In the left plot we show the relic density versus
the LSP mass.

The majority of points in \fig{fig:uni:random_relic} with $|\mu| \gtrsim 600$~GeV have a LSP with a small higgsino fraction, but two nearly degenerate neutralinos as lightest states, a pseudo-Dirac bino. 
As discussed in section \ref{SEC:MODEL}, the remaining spectrum is such that no s-channel resonances with sfermions in the propagator are possible. The reason is that Dirac gaugino masses do not contribute to the running of the sfermion mass terms. Hence, in this class of models  the
sfermion spectrum is more degenerate than in case of the MSSM and, thus, the LHC bounds on first and second generation squarks imply that all
of them are well above 1~TeV. Therefore,  the t-channel contribution to the dark matter annihilation is also suppressed, and  the mass difference to the LSP is too great, so that 
co-annihilation with sfermions
is by far not effective enough to reduce the relic density to an acceptable level.

\begin{figure}[t]
\centering
\includegraphics[width=.66\linewidth]{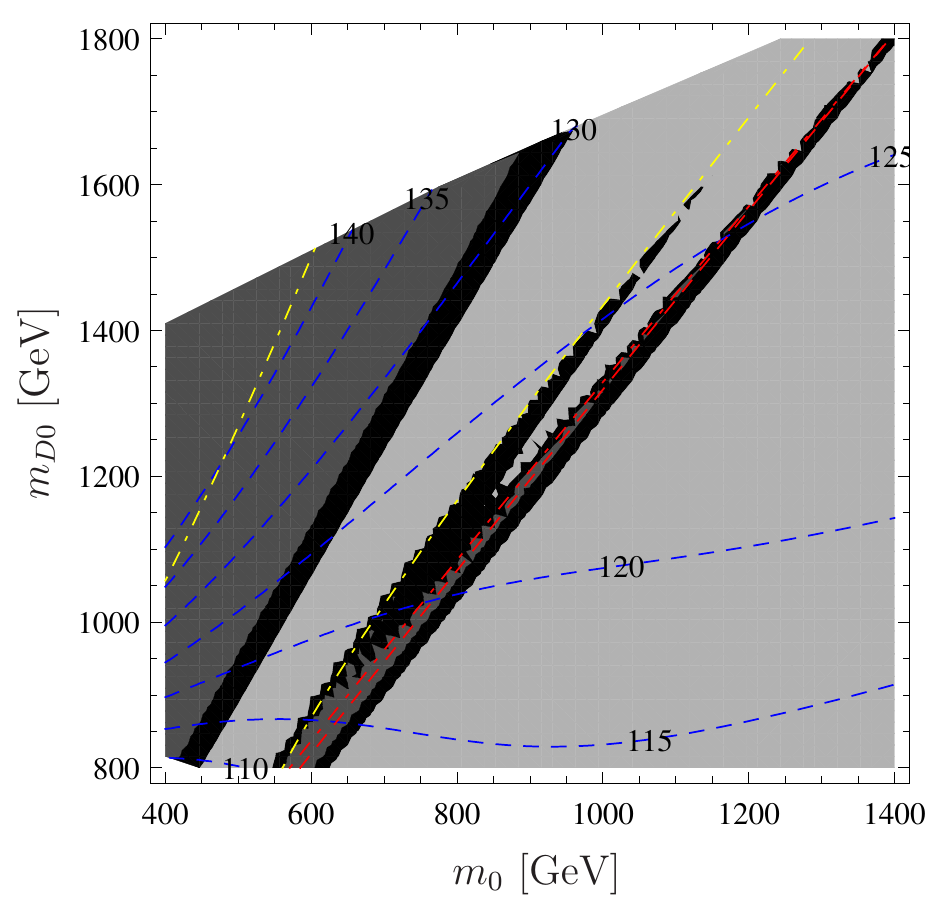}
\caption{Relic density in the $m_0 - m_{D0}$ plane using universal boundary conditions.
 The relevant parameters have been fixed to:
$\tan\beta=6$ and $B_{O}=-1.2 \cdot 10^6 {\rm{GeV}}^2$. 
Furthermore, we have chosen $\lambda_S=0.18$, 
$\lambda_T=0.39$
at the GUT scale which roughly corresponds to 
$\lambda_S=0.15$, 
$\lambda_T=0.52$
at the SUSY scale.
The blue dashed lines give the Higgs mass  in GeV.
The light (dark) grey area have a too large (too low) relic density whereas the
 black stripes correspond to parameter regions with the relic density within the bounds $0.05 < \Omega h^2 < 0.20$. The red (dot-) dashed lines indicate the rays  where the sum
of the two masses of the nearly degenerate, lightest neutralinos are equal to  the mass of the
(next-to) lightest pseudoscalar mass (Higgs funnel). The yellow dot-dashed lines indicate where the masses add up to the heavy scalar
Higgs mass. In the white, upper left region, the pseudoscalar octet becomes tachyonic despite $B_O \neq 0$ because of the large loop corrections due to  the heavy Dirac Gluino. 
}
\label{fig:uni:m0_mD_plane}
\end{figure}

The very narrow, mostly green and red strip at lower values for the relic density in the left plot 
of \fig{fig:uni:random_relic} (starting at ($m_{\tilde \chi_1^0} \simeq 250$~GeV, $\Omega h^2 \simeq 0.007$) and going to 
($m_{\tilde \chi_1^0} \simeq 1100$~GeV, $\Omega h^2 \simeq 0.1$)) is populated by 
points featuring an almost pure higgsino-like Dirac neutralino pair:
for parameter regions with small $m_0$ and comparatively large $m_{D0}$ we find several points
with the $|\mu| \lsim m_{1D}$, resulting in a higgsino-like LSP. 
This strip corresponds to the one with
$|N_{15}|^2+|N_{16}|^2
\approx 1$ in
the right plot.
This is analogous to the focus point region  in the CMSSM. 
As in the MSSM, pure
higgsinos annihilate too effectively and only at rather large masses of $m_{\tilde \chi^0_1} \approx |\mu| \gsim  800~$GeV
is the correct relic abundance obtained, in accordance with \eq{eq:analytics:higgsino_dm}.
For smaller values of $|\mu|$ a mixing with the bino-like states is necessary to get the correct relic density.

In addition we find several parameter points where the spectrum is such that  a 
$m_{\tilde \chi^0_1} \approx m_{A^0_i}/2$ or $m_{h^0_i}/2$,  i.e.\ an effective annihilation via
an $s$-channel resonance due to a heavy scalar or pseudoscalar Higgs boson is possible. This is the well-known
Higgs funnel which is also present in the CMSSM. However, due to the extended Higgs sector it does not only
occur for large but also for small $\tan\beta$.  We find that in general the
most likely final state is a $t \bar t$ pair in these scenarios.

There is no further mechanism for generating the correct range for $\Omega h^2$ in this version of the model,
because the sfermions are relatively heavy and thus, there is 
neither a bulk region nor is stau co-annihilation possible as already mentioned. 
Moreover, as we find a lower bound on the neutralino
mass of about 200~GeV, if all observations are to be explained simultaneously, neither the 
Higgs nor the $Z$ resonance can be present. 

%

In \fig{fig:uni:m0_mD_plane} we show the relic  density in the $m_0$-$m_{D0}$ plane fixing the remaining parameters
as indicated in the caption to exemplify our findings in more detail. Black regions have a relic density in the range
$0.05 < \Omega h^2 < 0.2$ whereas the light (dark)  grey have values below (above) this range.
In the upper left white area the pseudoscalar octet becomes tachyonic.
Apart from the obvious focus point region which extends in the area with 
low $m_0$ and large $m_{D0}$, one can see the distinct Higgs 
funnel which consists of two separate annihilation strips: on the right, the (co)annihilation  
$\tilde \chi^0_i \tilde \chi^0_j \to A^0 \to t \bar t$, $i,j=1,2$ causes the small relic abundance, on the
left the intermediate heavy scalar Higgs is the reason for the (slightly less pronounced) funnel; as discussed in \ref{SEC:ANALYTIC}, these two states are almost degenerate. 
For completeness we give also the value of the SM-like Higgs mass (dashed lines).

\begin{figure}[t]
\centering
\includegraphics[width=.49\linewidth]{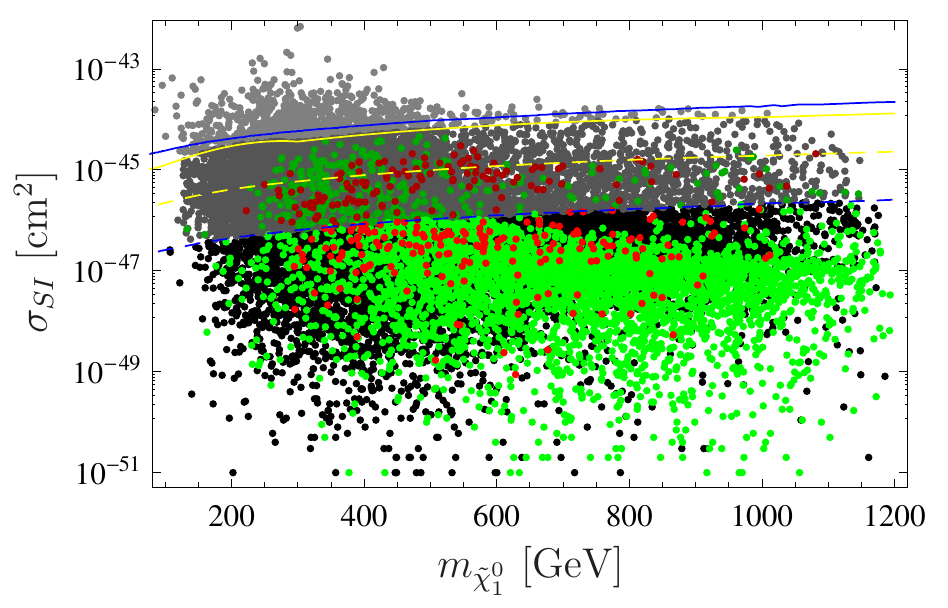}
\includegraphics[width=.49\linewidth]{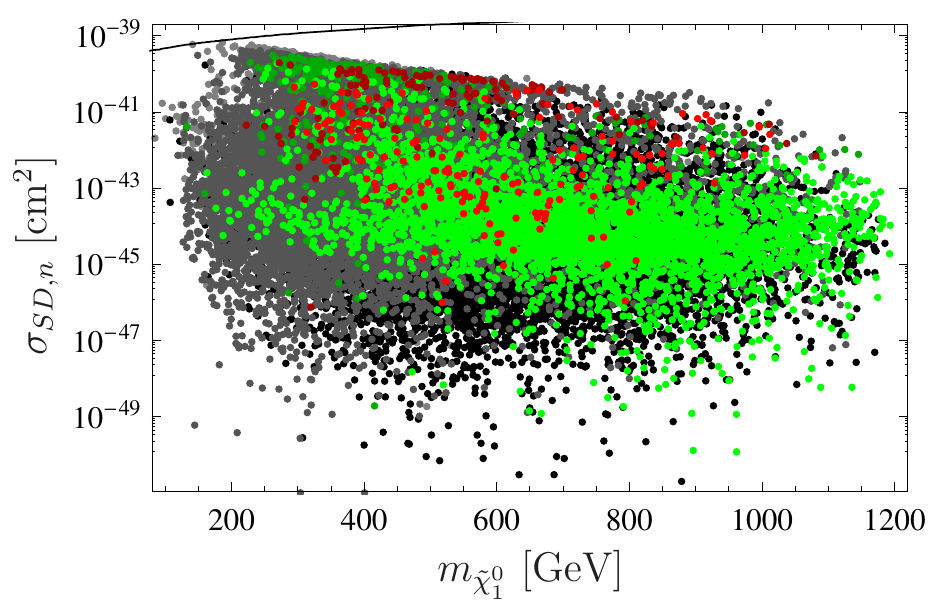}
\caption{Cross section of the spin-independent (left side)  cross section
of the dark matter candidate with neutrons. The parameter points and the colour coding are the
same as in 
\fig{fig:uni:random_relic}. 
The full blue and yellow lines show the current bounds on $\sigma_{SI}$
from XENON100 and LUX, respectively. The projections for the LUX run 2013/14 (yellow dashed line)
and  XENON1T (blue dashed line) are  shown as well. 
The  corresponding spin-dependent  cross section is given in the right plot where the
black line gives
the current upper limit on spin-dependent annihilation cross sections $\sigma_{SD,n}$ as from XENON measurements.
}
\label{fig:uni:random_DD}
\end{figure}

\Fig{fig:uni:random_DD} presents the neutralino-nucleon cross sections vs.\
 the lightest neutralino mass, both for spin-independent  and 
spin-dependent  measurements. While the present-day spin-independent measurements are not sensitive to the scenarios under consideration,
the spin-dependent ones already cut 
into the parameter space, mostly for light (i.e. $m_{\tilde \chi^0_1}<500~$GeV) higgsino-like LSPs. 
We also show  the current LUX \cite{Akerib:2013tjd} and
 XENON100 \cite{Aprile:2012nq, Aprile:2013doa}  bounds, and
the projections of the exclusion potential of the 2013/2014 LUX run \cite{Gaitskell:2013} 
and the XENON upgrade \cite{XENON:2013}. Clearly for a significant part  of the
parameter space with a mixed bino/higgsino LSP
either a signal can be detected or otherwise the corresponding parameters can be excluded.
However, nearly pure higgsino-like LSPs are hardly covered,  i.e.\ they will be neither discovered nor 
excluded by these experiments.


\subsubsection{Case 2: non-zero $M_S$}
\label{subsubsec:MSneqZero}
%
\begin{figure}[t]
\centering
\includegraphics[width=.49\linewidth]{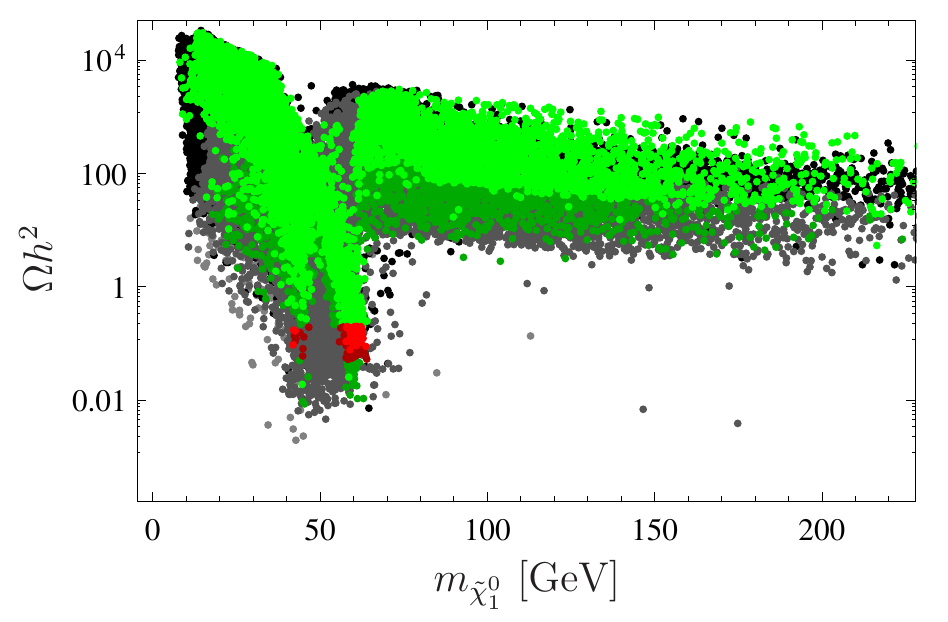}
\includegraphics[width=.49\linewidth]{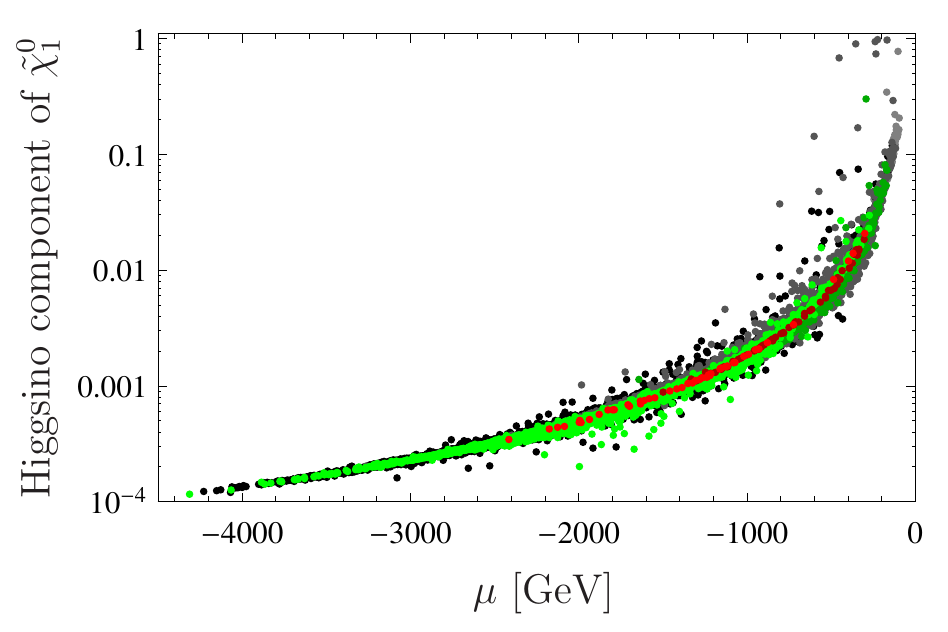}
\caption{Random scan over the model parameter space using universal boundaries and $10 > M_S/{\rm{TeV}} > 0$. The colour code is as in the previous figures. 
The figure on the left panel shows the relic density vs. the mass of the LSP. The two dips in $\Omega h^2$ correspond to parameter points where $m_{\tilde \chi^0_1} = m_{Z}/2$ and $m_h/2$.
The right figure shows the higgsino content vs. the $\mu$ parameter.
}
\label{fig:uni:random_relic_MSneq0}
\end{figure}
We have seen so far that in the universal case only two possibilities exist to find the correct
relic density: either a moderate-to-large higgsino fraction of the LSPs
or a Higgs funnel with either a heavy scalar  or pseudoscalar Higgs resonance. Other possibilities are highly constrained because 
the parameters in the neutralino mass matrix also have  a large impact on other 
aspects of the model. A lighter higgsino-like LSP is forbidden by the minimum conditions of the vacuum, and lighter bino-singlino
like states are ruled out because of the strong relation to the Gluino mass. One can try to circumvent these interplays 
by allowing for additional parameters in the neutralino mass matrix. The simplest option is to allow for $M_S\ne 0$. 
This term breaks $R$-symmetry, and is difficult to motivate from a top-down approach. However, 
it does not introduce Majorana mass terms for gauginos nor trilinear soft-breaking 
couplings for sfermions. Thus, the most interesting, phenomenological relevant differences compared to the (C)MSSM are kept. 
In addition, the impact of $M_S$ on the properties of other particles, in particular the Higgs scalars, is very moderate.

\begin{figure}[t]
\centering
\includegraphics[width=.49\linewidth]{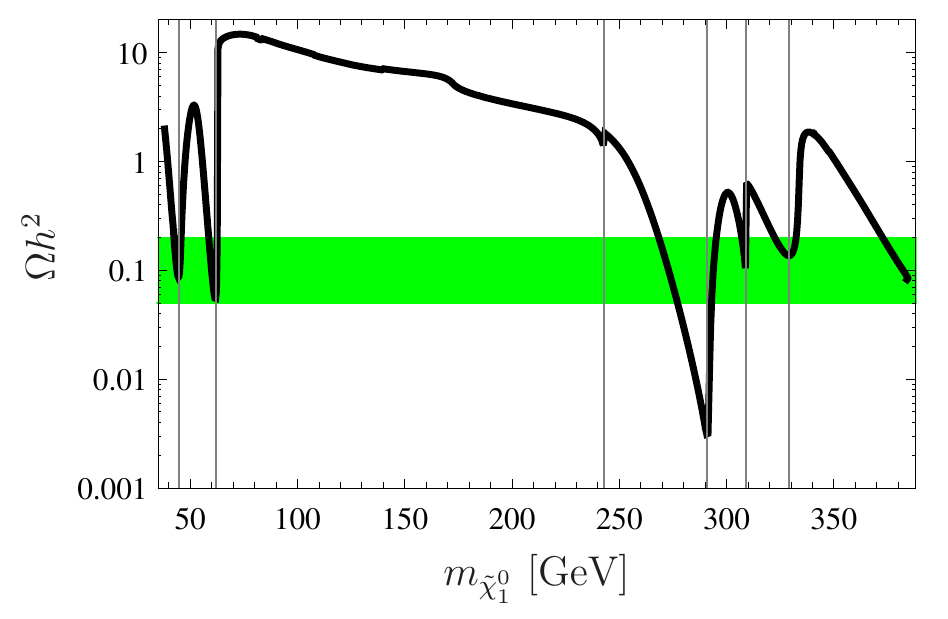}
\caption{The neutralino relic density as function of the LSP mass for a variation of $M_S$ in the range  $M_S =[0,5.5]~\rm{TeV}$ using universal boundary conditions. 
$M_S = 0$ corresponds to the right end of the plot whereas larger values
lead to a suppression of the neutralino mass. The other parameters have been chosen as in \fig{fig:uni:m0_mD_plane} with fixed $m_0 = 0.6~$TeV and $m_{D0}=1.1~$TeV.
  The narrow green band shows the measured relic density within 
the experimental error. 
The exact positions of the  resonances are indicated by vertical lines. They correspond to the annihilation via $X=Z,h,A^0_1,A^0_2,A^0_3,h_2$ (from left to right)
}
\label{fig:uni:MS_scans}
\end{figure}

We have already seen in  the discussion 
of the neutralino mass matrix in \SEC{sec:model:neutralinos}, that the main effect of non-vanishing
$M_S$ consists of splitting
the bino/singlino Dirac state into two Majorana particles even without the mixing with the higgsinos. 
The mass of the lighter state is roughly
 $m_{1D}^2/M_S$ if $M_S \gg m_{1D}$.  
Thus, depending on the mass ratio $m_{1D}/M_S$, the resulting neutralino state can
become, in principle, arbitrarily light. Note that this state is a nearly pure gauge singlet with only
a very small higgsino  contribution as can also be seen on the 
the right plot of \fig{fig:uni:random_relic_MSneq0}. Therefore, even neutralino masses $O(10~\text{GeV})$
are not in conflict with data.
In the left plot of  this figure
we show the results in the $\Omega h^2 - m_{\tilde \chi^0_1}$ plane using our standard colour 
coding. Due to the small higgsino content
new mechanisms are needed to obtain the proper relic density: we find resonances with the $Z$-boson or
 the light Higgs boson are viable possibilities.
The two dips in this plot correspond to
$m_{\rm{LSP}} \simeq  m_{Z}/2$ and $m_{\rm{LSP}} \simeq  m_{h}/2$. However, as the first possibility 
requires the admixture of a higgsino, the corresponding dip in $\Omega h^2$ is not as 
pronounced as the corresponding Higgs-mediated one. In \fig{fig:uni:MS_scans} 
we show the relic density as a function
of $m_{\tilde \chi^0_1}$. Here we have taken the same parameters as for \fig{fig:uni:m0_mD_plane}
together with $m_0 = 0.6~$TeV and $m_{D0}=1.1~$TeV. The variation of $m_{\tilde \chi^0_1}$ stems from
the variation $0 < M_S/{\rm{TeV}} < 5.5$. The case $M_S=0$ correspond to the largest neutralino
mass  and and the LSP for this region has a sizeable higgsino admixture of roughly 65~\%  allowing 
an effective  
$\tilde \chi^0_i \tilde \chi^0_j \to A B$ ($i,j=1,2$) annihilation.
 With increasing $M_S$, this admixture is quickly suppressed
and is already below 2~\%  for $m_{\tilde \chi^0_1} \lsim 350$~GeV yielding an overabundant relic density.
The rest of the plot is governed by spikes of which each implies a different resonant annihilation channel
$\tilde \chi^0_1 \tilde \chi^0_1 \to X  \to A B$, i.e.\ the peaks occur at $m_{\tilde \chi^0_1} \simeq m_X/2$ and correspond
with decreasing $m_{\tilde \chi^0_1}$ to $X=h_2,A^0_3,A^0_2,A^0_1,h,Z$. 
 The notch close to $m_{\tilde \chi^0_1} \approx 173~$GeV stems from
the opening of the $t \bar t $ final state. 
We want to stress that \fig{fig:uni:MS_scans} features every single mechanism for generating 
the observed relic abundance that is possible within this model for universal scalar masses at the GUT scale.

%
\begin{figure}[t]
\centering
\includegraphics[width=.49\linewidth]{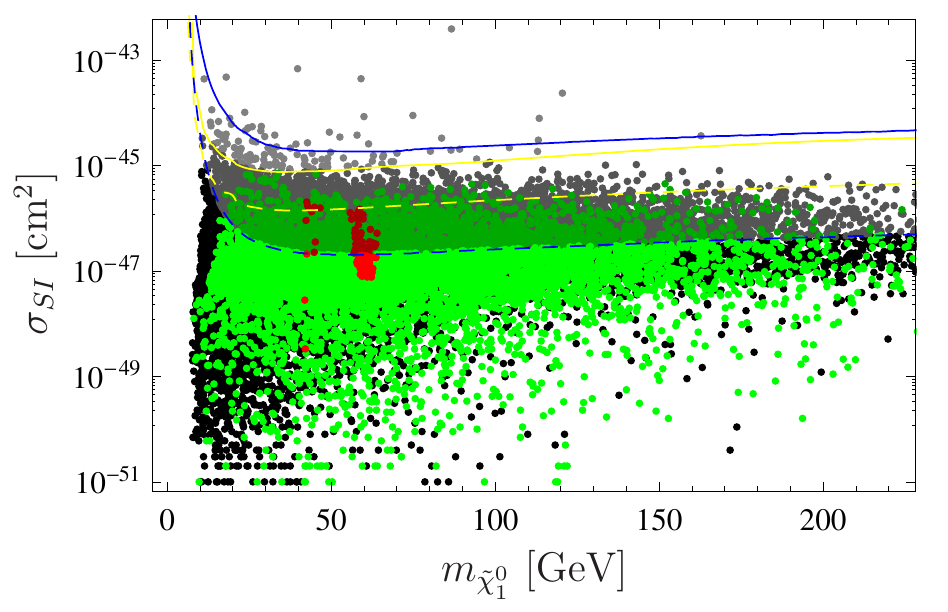}
\includegraphics[width=.49\linewidth]{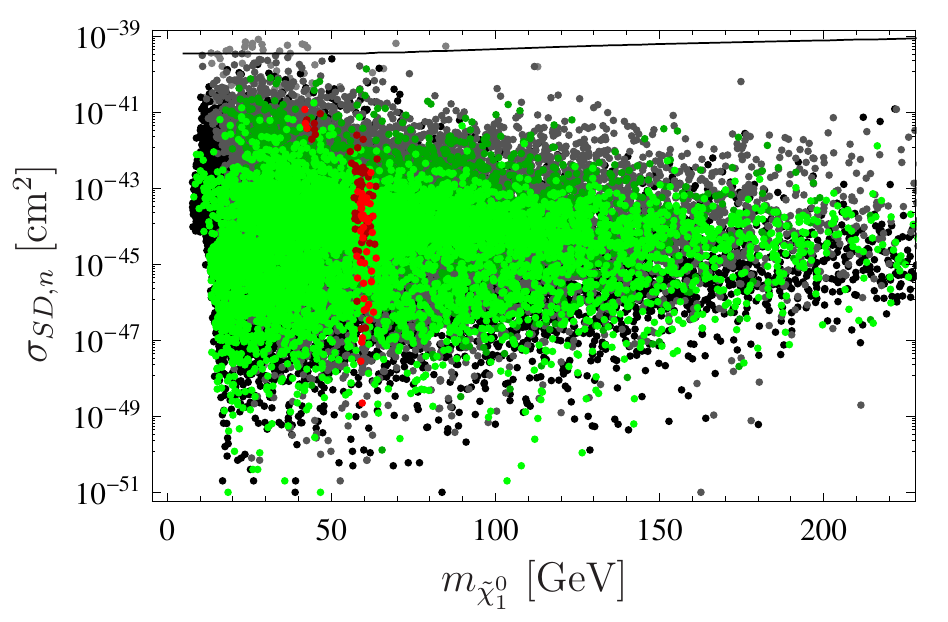}
\caption{Analogue to \fig{fig:uni:random_DD} for the case of non-zero $M_S$.}
\label{fig:uni:random_DD_MSneq0}
\end{figure}

The neutralino-nucleon cross sections are generically lower than
in the former case without $M_S$, see figure~\ref{fig:uni:random_DD_MSneq0}, 
because of the dominating bino/singlino nature of the DM candidate. 
As a consequence, almost none of the possible configurations are excluded 
by current direct detection experiments.
 However, the next generation of experiments will be able to probe a significant portion of the parameter space allowing for a Higgs resonance: 
most scenarios with a higgsino admixture of roughly two per mille or more (or, conversely, scenarios with $|\mu| < 1~$TeV)  
have cross-sections within the reach of XENON1T, cf. \fig{fig:uni:random_relic_MSneq0}.


\subsection{Non-universal boundary conditions}

\begin{table}[t]
\centering
\begin{tabular}{l|r c l||l|r c l} \hline \hline
 ~~$m_0$~[GeV]~~ & 1000 &...& 6000 & ~~$m_\Sigma$~[GeV]~~ & 1200 &...& 4200 \\
 ~~$m_{D0}$~[GeV]~~ & 500 &...& 1700 & ~~$m_S$~[GeV]~~ & $100$ &...& 1400 \\
 ~~$\tan \beta$~~ & 1.5 &...& 3 & ~~$\mu$~[GeV]~~ & $-1000$ &...& $-150$ \\
 ~~$\lambda_S^2 + 2 \lambda_T^2$~~ & $0.45^2$ &...& $0.75^2$ & ~~$\sqrt{B_\mu}~$[GeV]~~ & 200 &...& 1200 \\
 \hline 
 \hline
\end{tabular}
\caption{Ranges of the varied parameters in the non-universal Higgs mass scenario. 
All values are GUT-scale input, except for $m_S$, $\lambda_S$ and $\lambda_T$ which are defined
at the electroweak scale.  $M_S$, $B_S$ and $B_O$ are assumed to be zero.}
\label{tab:nonuni:parameter_ranges}
\end{table}

We shall now relax the boundary conditions at the GUT scale by allowing soft terms for 
the Higgs doublets which are not identical to those of the other scalars. This can be motivated by assuming 
an underlying GUT theory such as $SO(10)$ where all matter fields come from three generations of  a {\bf 16}, but Higgs fields 
descend from other representations. Unification of the gauge groups should unify the masses of the triplet and octet adjoint scalars, but depending on the embedding the singlet adjoint may have a different value (as considered in \cite{Benakli:2014cia}). Finally, 
we trade the input values of $m_{H_d}^2$ and $m_{H_u}^2$ for $\mu$ and $B_\mu$, i.e. the $\mu$ and $B_\mu$ are free parameters, 
while the soft-terms for the doublets are obtained from the tadpole conditions. 

The parameter ranges considered are given in \TAB{tab:nonuni:parameter_ranges}. Note that we have chosen
$0.45^2 < \lambda_S^2+2\lambda_T^2 < 0.75^2$ at the SUSY scale as this allows us better control of
the mass of the lighter scalar Higgs boson \cite{Benakli:2012cy}. 

\begin{figure}[t]
\centering
\includegraphics[width=.49\linewidth]{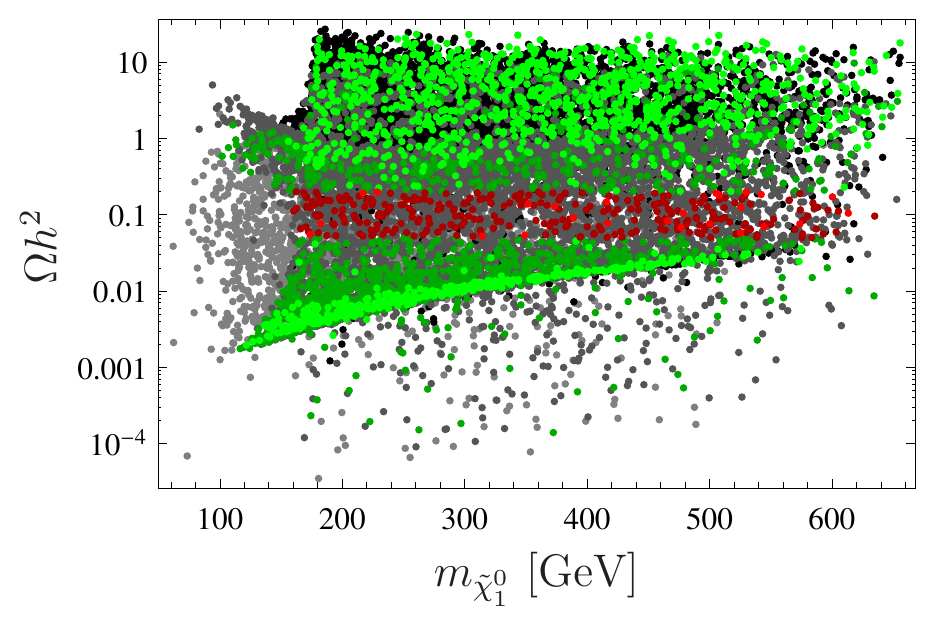}
\caption{Relic density $\Omega h^2$ versus $m_{\tilde \chi^0_1}$  using non-universal Higgs masses and the parameter 
ranges of  \TAB{tab:nonuni:parameter_ranges}. The colour code is as in \fig{fig:uni:random_relic}.}
\label{fig:nonuni:random_relic}
\end{figure}

\begin{figure}[t]
\centering
\includegraphics[width=.49\linewidth]{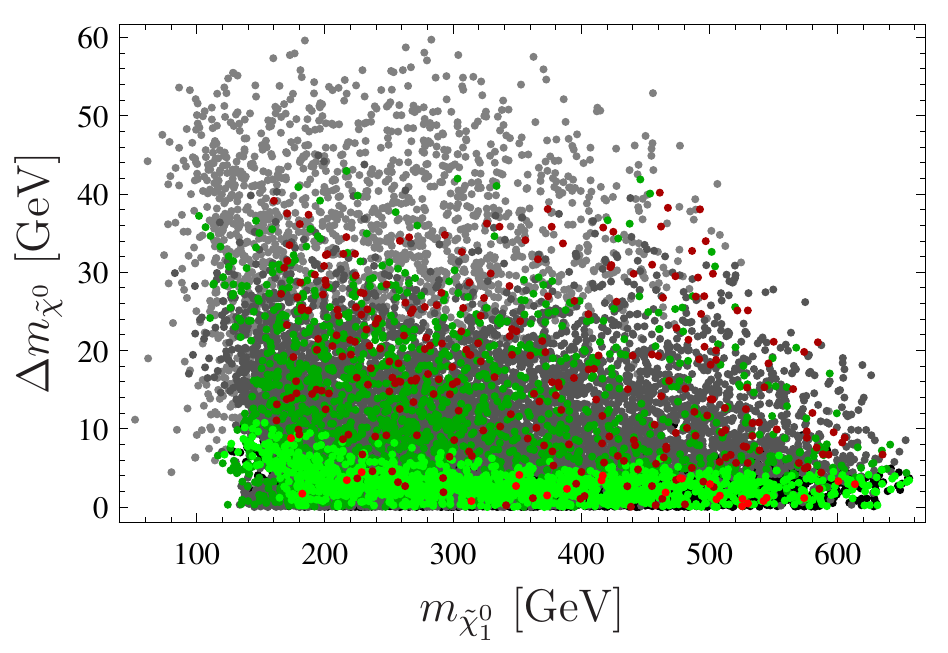}
\includegraphics[width=.49\linewidth]{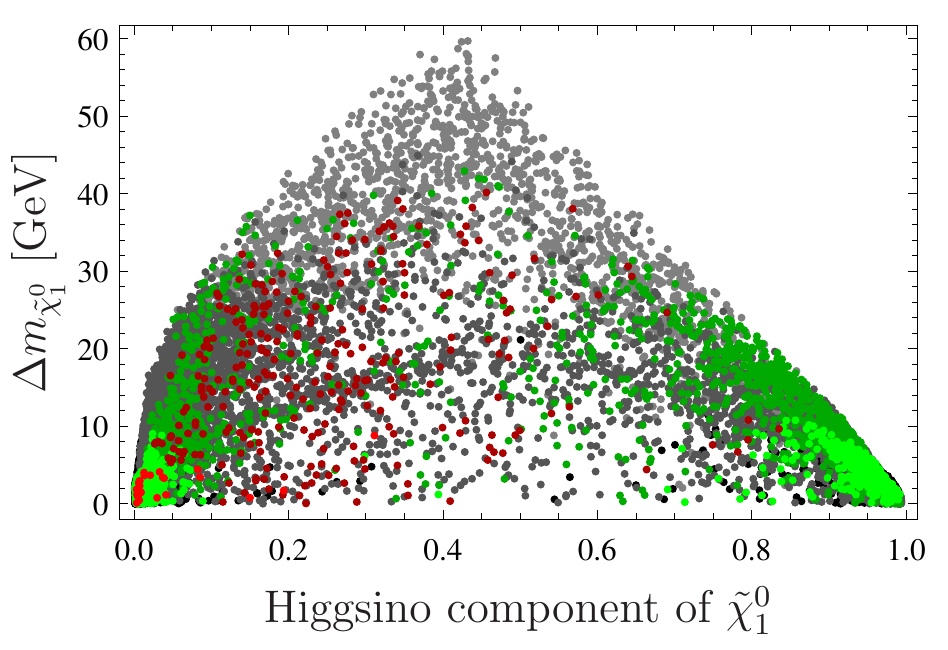}
\caption{$\Delta m_{\tilde \chi^0} = m_{\tilde \chi^0_2}-m_{\tilde \chi^0_1}$ for the same points as in
\fig{fig:nonuni:random_relic} using the same colour code.
All points with right relic density  have $\Delta m_{\tilde \chi^0} \leq$ 40 GeV. Parameter points which cannot be excluded by XENON1T all have even smaller mass differences. All points with a small higgsino fraction ($\lesssim 0.05$) that satisfy the correct density lie very close to a resonance. 
}
\label{fig:nonuni:random_dmChi}
\end{figure}

In \fig{fig:nonuni:random_relic} we show  $\Omega h^2$ versus $m_{\tilde \chi_1^0}$ for a random scan
over these parameter ranges. At first glance we find similar features as in the case of universal boundary
conditions (UBC). However, we find  an upper limit of $m_{\tilde \chi_1^0} \lsim$~700~GeV.
This can be understood as follows:
the large values of $\lambda_S^2+2\lambda_T^2$ at the SUSY scale which are required for the mass of the lighter
Higgs to be close to 125 GeV increase during the RGE evolution to the GUT scale. 
This in turn implies that the
$\mu$-parameter has also a significant RGE evolution \footnote{This is in contrast to the CMSSM where it is
only changed by a few per-cent}) resulting in a significantly lower value for $\mu$ at the SUSY scale, yielding
the observed upper bound on $m_{\tilde \chi_1^0}$.
As a consequence the higgsino content of $\tilde \chi_1^0$ is even for dominantly bino/singlino-like LSPs
larger than in the UBC case. This is the reason why the maximal relic density found is an order
of magnitude smaller compared to the UBC case.
In \fig{fig:nonuni:random_dmChi} 
 we give the mass splitting between the lightest neutralinos versus $m_{\tilde \chi^0_1}$ (right plot)
and the higgsino content (left plot).
 We find that the mass splitting $\Delta \tilde \chi^0$
between the lightest and second lightest neutralino  to be smaller than 60 GeV for all parameter points and smaller than 40 GeV
for points with the correct dark matter properties, demonstrating that co-annihilation plays a large role accross the whole parameter space.
Finally, 
we find an lower bound on $m_{\tilde \chi_1^0}$ of about 200~GeV if all observations should
be explained simultaneously which is similar to the UBC case.

\begin{figure}[t]
\centering
\includegraphics[width=0.49\linewidth]{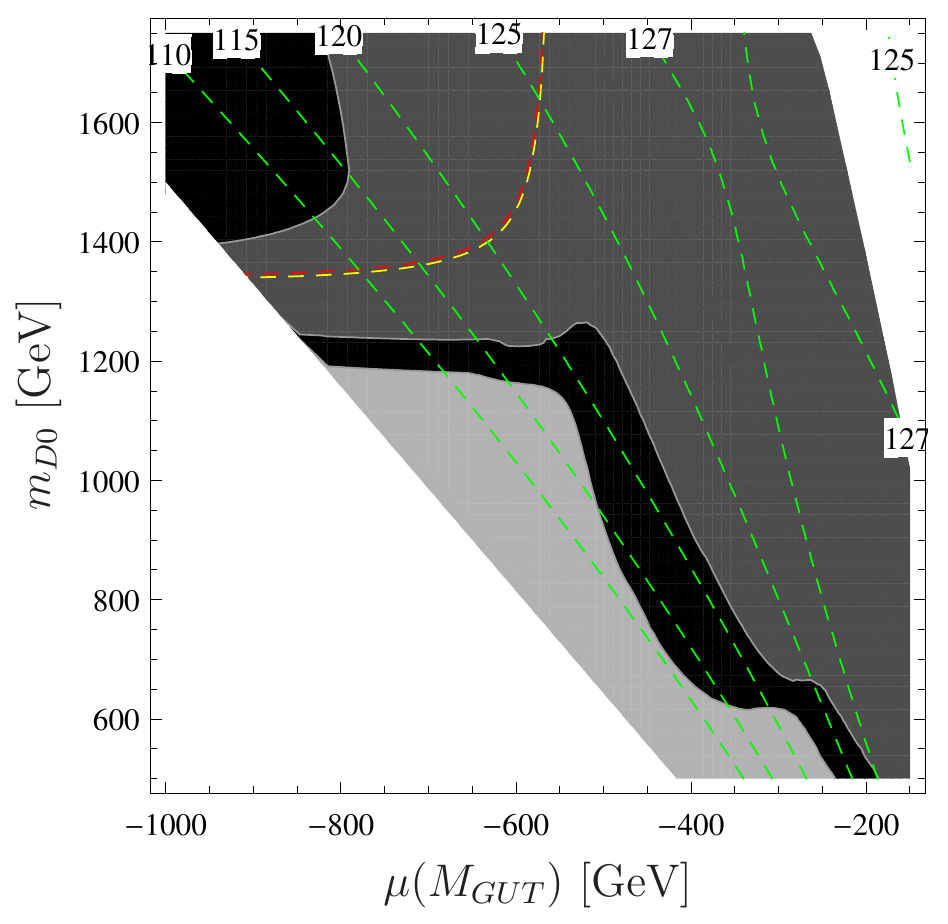}
\includegraphics[width=0.49\linewidth]{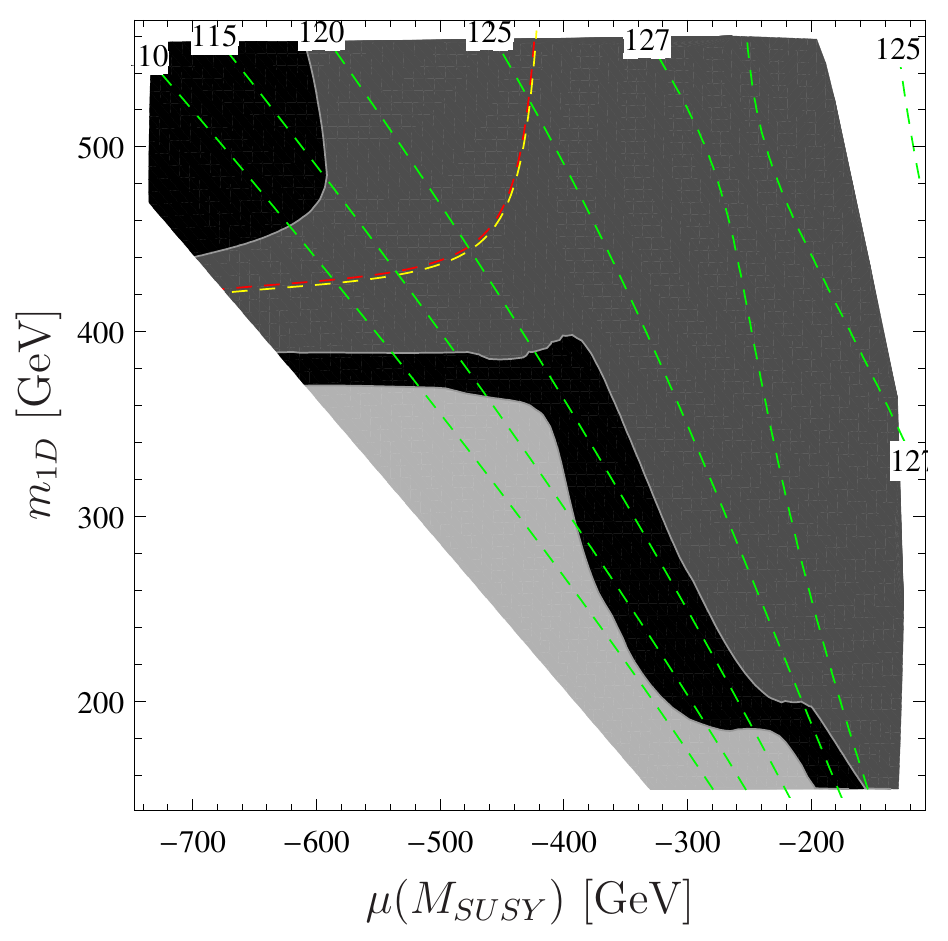}
\caption{Contours of the relic density in the $\mu$-$m_{D0}$-plane for
 $m_0$ = 4.1 TeV, $\tan\beta$ = 2.71, $B_\mu = 4.7 \cdot 10^5$ GeV$^2$, $m_S$ = 370 GeV, $m_\Sigma$ = 3.4 TeV, $\lambda_S$ = 0.5 and 
 $\lambda_T = 0.37$.
Black regions have a relic density within $0.05<\Omega h^2 <0.2$, in dark (light) grey regions the relic density is too low (high). Green dashed lines give the mass of the lightest Higgs boson, i.e.\ the one resembling the SM-Higgs boson. 
On the red (yellow) line one has $m_{A_2} = 2 m_{\tilde \chi^0_1}$ ($m_{H_2} = 2 m_{\tilde \chi^0_1}$).
The upper white corner is excluded due to a chargino LSP and the lower right one due to a tachyonic Higgs state. In the left figure parameters are shown at the GUT scale, in the right one at SUSY scale. }
\label{fig:nonuni:mu_mD_plane2}
\end{figure}

In \fig{fig:nonuni:mu_mD_plane2} we show the relic density in the $\mu$-$m_{D0}$-plane, analogous to the $m_0$-$M_{1/2}$-plane in 
the CMSSM, where the remaining free parameters were fixed to the values given in the caption. 
In the light (dark) grey coloured regions the relic density is too high (low), whereas
in the black areas we find a relic density of $0.05 < \Omega h^2 < 0.2$. The upper left white corner is
excluded due to a chargino LSP and in the bottom right one the lightest Higgs becomes tachyonic. 
In the lower black region one has co-annihilation between the higgsino-like and the bino/singlino like states. 
$m_{D0} \approx 600~\rm{GeV}$ corresponds to  $m_{\tilde \chi_1^0} \approx 173 \rm{GeV}$ and, thus for this and larger
values the $t\bar{t}$ final state opens up which is the reason for the shift  and broadening of the black band. Indeed the region with the correct relic density for $-400\ \mathrm{GeV} \lesssim \mu \lesssim -250$ GeV can be at least partly described by equation (\ref{EQ:higgsportalomega}).

The yellow (red) line indicates where the LSP has half the mass of the second lightest (pseudo-)scalar
Higgs boson corresponding to the Higgs funnel which actually lead to an underabundant relic density here.
The black upper left corner is a combination of co-annihilation and the effects of a not-too-off-shell Higgs state.

\begin{figure}[t]
\centering
\includegraphics[width=.49\linewidth]{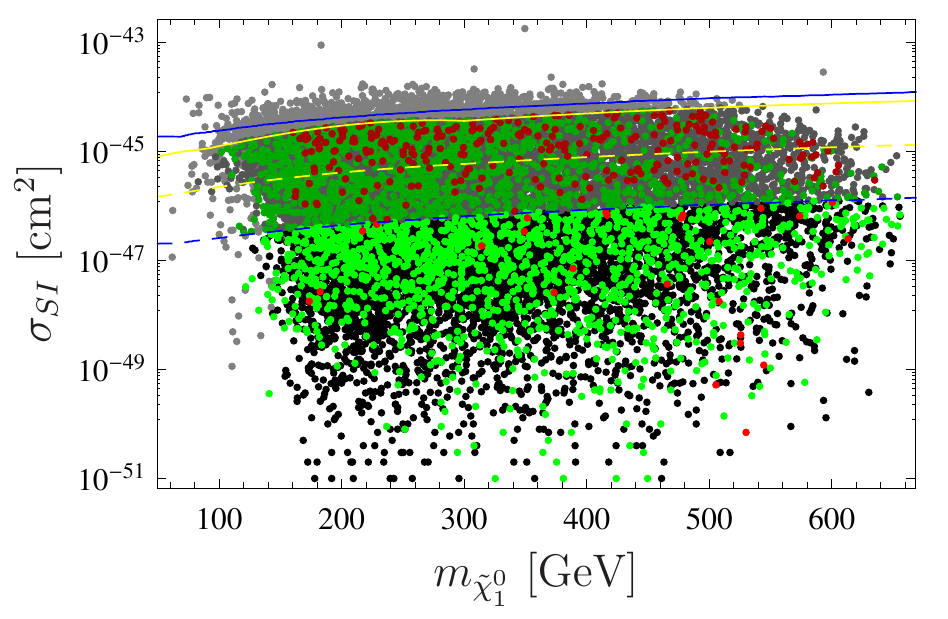}
\includegraphics[width=.49\linewidth]{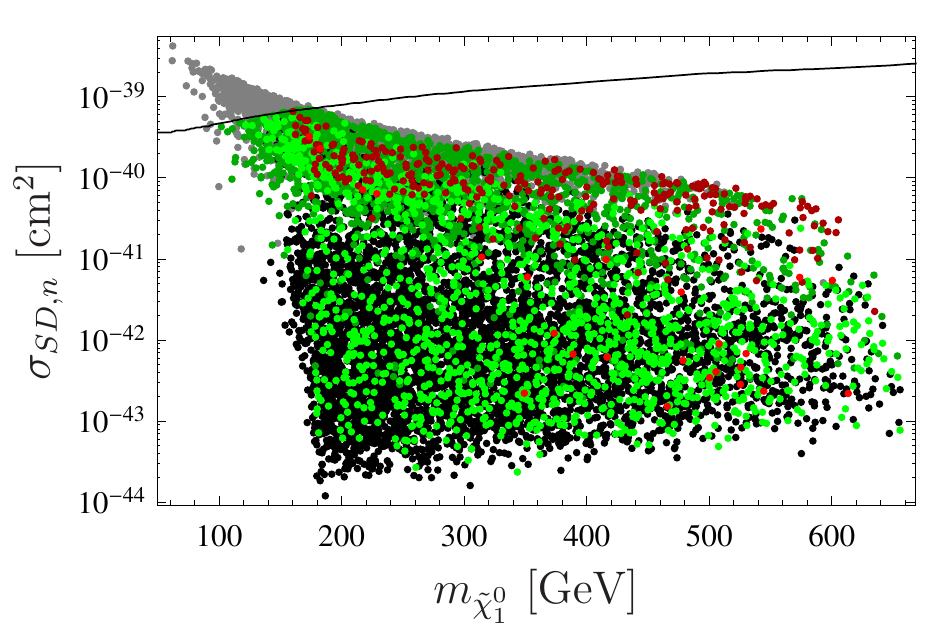}
\caption{Spin-independent annihilation cross section $\sigma_{SI}$ and spin dependent annihilation cross-section
 with neutrons
 $\sigma_{SD,n}$ of the dark matter candidate for the data from figure \ref{fig:nonuni:random_relic} using the same colour code. The blue and yellow lines show the current bounds on $\sigma_{SI}$
from XENON100 and LUX respectively. Dashed lines are projections for the LUX run 2013/14 and the currently under construction XENON1T. Almost all points with correct Higgs mass and relic density can therefore be probed with the next generation of detectors.
The black line in the right figure shows the XENON100 bounds on $\sigma_{SI,n}$. All data points fulfil the XENON100 bounds on neutralino-proton cross sections.}
\label{fig:nonuni:random_DD}
\end{figure}

To finish, we show in \fig{fig:nonuni:random_DD}  the neutralino-nucleon cross sections of the same parameter points 
of \fig{fig:nonuni:random_relic} together with the current experimental 
limits set by LUX and XENON100. The results are again similar to the UBC case but with one important difference:
nearly all points which are consistent with existing data,  that means within the extended range due to unknown
theoretical uncertainties, will be probed by XENON1T. This is again a consequence of the increased higgsino-content
of the LSP.


\section{Conclusions}
\label{SEC:CONCLUSIONS}
We have investigated possible dark matter scenarios in the CMDGSSM. In this model, the pure gauginos are Dirac states. Only due to the mixing with the higgsinos, which are Majorana particles because of $R$-symmetry breaking, the lightest neutralino is always a Majorana DM candidate with a small mass splitting to the next lightest state. Due to the heavy spectrum of superpartners the annihilation is generally dominated by the exchange of a SM-like Higgs scalar or $Z$-boson.

We studied the case of fully universal scalar masses at the GUT scale and the case with non-unified masses for the Higgs doublets and adjoints. In the minimal version of the first case, with  only $B_\mu$ as an $R$-symmetry breaking parameter in the potential, we found a lower mass of the LSP with correct relic density of about 200~GeV. 
For this mass range, viable dark matter candidates are nearly degenerate bino/singlino states
whereas the most efficient annihilation mechanism to suppress the relic density to the allowed level is via resonances with heavy scalars and pseudoscalars. Higgsino-like neutralinos with the correct relic density
are only found for LSP masses of about 800~GeV as per the classic formula when other states are decoupled. 

If we allow for $M_S$ to be non-vanishing, lighter neutralinos could comprise the observed DM of the universe. On the other hand, if we surrender the condition of complete unification in the soft-breaking scalar sector, light higgsinos can more easily be obtained and so the higgsino admixture of a bino/singlino pair is in general larger. As a consequence, co-annihilation between the bino LSP and higgsino-like particles is possible. 

For all  cases considered we found that the current direct detection experiments are only slightly constraining. However, the next generation of experiments like XENON1T or the next run of LUX can probe a large part of the parameter space with higgsino-like DM candidates and $|\mu| < 1$~TeV. Naturally, the bino/singlino option is more difficult to test experimentally. Therefore, the scenario with non-vanishing $M_S$ and a light LSP would still not be addressed by these experiments.

\section*{ACKNOWLEDGEMENTS}
We would like to thank Karim Benakli for numerous interesting discussions throughout this work
and Timon Emken for his input in the early stage of this project.
MEK and WP are supported by the DFG research training group
GRK1147 and by the DFG project no. PO-1337/3-1.



\end{document}